\documentclass[sigconf]{acmart}
\usepackage{array}
\usepackage{subcaption} 
\usepackage{multirow}
\usepackage{makecell}
\usepackage{fontawesome}
\usepackage{float}

\AtBeginDocument{%
  }

\setcopyright{acmlicensed}
\copyrightyear{2025}
\acmYear{2025}

\acmConference[CHI '25]{CHI Conference on Human Factors in Computing Systems}{April 26-May 1, 2025}{Yokohama, Japan}
\acmBooktitle{CHI Conference on Human Factors in Computing Systems (CHI '25), April 26-May 1, 2025, Yokohama, Japan}
\acmDOI{10.1145/3706598.3714159}
\acmISBN{979-8-4007-1394-1/25/04}

\definecolor{vividauburn}{rgb}{0.58, 0.15, 0.14}
\definecolor{rufous}{rgb}{0.66, 0.11, 0.03}
\definecolor{red-brown}{rgb}{0.65, 0.16, 0.16}
\definecolor{red(ncs)}{rgb}{0.77, 0.01, 0.2}
\newcommand{\revisedtext}[0]{}
\newcommand{\rrtext}[0]{}

\begin{document}

\title{HarmonyCut: Supporting Creative Chinese Paper-cutting Design with Form and Connotation Harmony}

\author{Huanchen Wang}
\orcid{0000-0001-9339-1941}
\authornote{Equal contribution.}
\affiliation{
  \department{Department of Computer Science and Engineering}
  \institution{Southern University of Science and Technology}
  \city{Shenzhen}
  \state{Guangdong}
  \country{China}
}
\affiliation{
  \department{Department of Computer Science}
  \institution{City University of Hong Kong}
  \city{Hong Kong}
  \country{China}
}
\email{wanghc2022@mail.sustech.edu.cn}

\author{Tianrun Qiu}
\orcid{0009-0002-9878-663X}
\authornotemark[1]
\affiliation{
  \department{Department of Computer Science and Engineering}
  \institution{Southern University of Science and Technology}
  \city{Shenzhen}
  \state{Guangdong}
  \country{China}
}
\email{qiutr@mail.sustech.edu.cn}

\author{Jiaping Li}
\orcid{0000-0003-0128-8993}
\affiliation{
  \department{Department of Computer Science and Engineering}
  \institution{Southern University of Science and Technology}
  \city{Shenzhen}
  \state{Guangdong}
  \country{China}
}
\email{lijp2024@mail.sustech.edu.cn}

\author{Zhicong Lu}
\orcid{0000-0002-7761-6351}
\affiliation{
  \department{Department of Computer Science}
  \institution{George Mason University}
  \city{Fairfax}
  \state{Virginia}
  \country{USA}
}
\email{zlu6@gmu.edu}

\author{Yuxin Ma}
\authornote{Corresponding author.}
\orcid{0000-0003-0484-668X}
\affiliation{
  \department{Department of Computer Science and Engineering}
  \institution{Southern University of Science and Technology}
  \city{Shenzhen}
  \state{Guangdong}
  \country{China}
}
\email{mayx@sustech.edu.cn}


\begin{abstract}
Chinese paper-cutting, an Intangible Cultural Heritage (ICH), faces challenges from the erosion of traditional culture due to the prevalence of realism alongside limited public access to cultural elements. While generative AI can enhance paper-cutting design with its extensive knowledge base and efficient production capabilities, it often struggles to align content with cultural meaning due to users' and models' lack of comprehensive paper-cutting knowledge. To address these issues, we conducted a formative study (N=7) to identify the workflow and design space, including four core factors (Function, Subject Matter, Style, and Method of Expression) and a key element (Pattern). We then developed HarmonyCut, a generative AI-based tool that translates abstract intentions into creative and structured ideas. This tool facilitates the exploration of suggested related content (knowledge, works, and patterns), enabling users to select, combine, and adjust elements for creative paper-cutting design. A user study (N=16) and an expert evaluation (N=3) demonstrated that HarmonyCut effectively provided relevant knowledge, aiding the ideation of diverse paper-cutting designs and maintaining design quality within the design space to ensure alignment between form and cultural connotation. 

\end{abstract}

\begin{teaserfigure}
  \includegraphics[width=\textwidth]{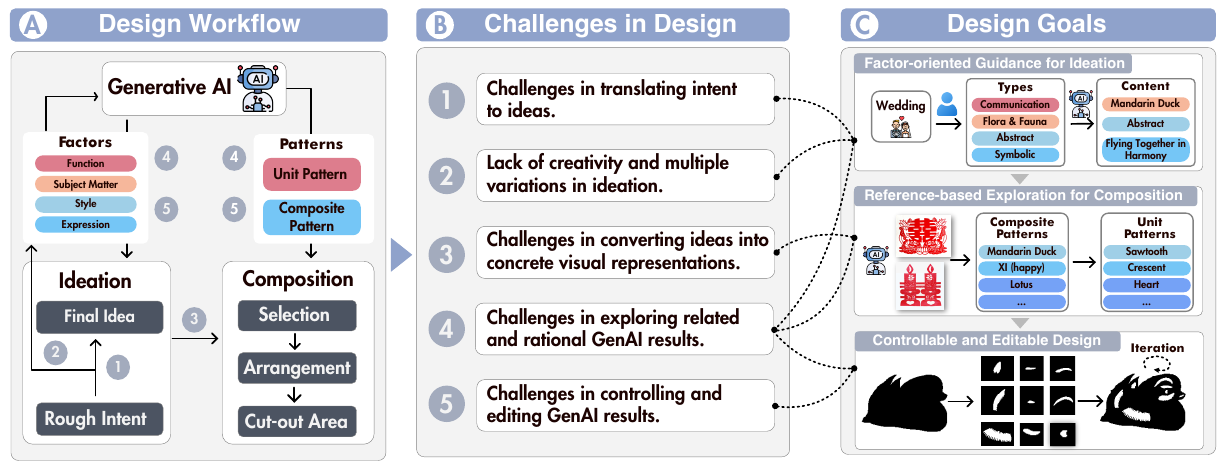}
\caption{\label{figure1}
A general two-stage (ideation and composition) workflow and design space (four factors and one element) for GenAI-aided paper-cutting design is outlined from the two-step formative study, with the main challenges in the workflow labeled on corresponding stages. Based on the workflow and challenges, the design goals are solidified in the pipeline and interface of HarmonyCut.}
\Description{This figure shows a general two-stage (ideation and composition) workflow and design space (four factors and one element) for GenAI-aided paper-cutting design outlined from the two-step formative study, with the main challenges in the workflow labeled on corresponding stages. Based on the workflow and challenges, the design goals are solidified in the pipeline and interface of HarmonyCut.}
  \label{fig:teaser}
\end{teaserfigure}


\maketitle

\section{Introduction}
\revisedtext{Paper-cutting, widely recognized as an artistic form across many countries, stands as a powerful medium for cultural exchange and a bridge for global connections. In the context of contemporary globalization, the preservation and inheritance of this traditional craft is crucial, as its distinctive form continues to foster cultural understanding and advance cross-cultural communication~\cite{Zhang:2000:internationalpapaercut, Shu:2023:cross-culturalpapercut}.}
Within this broad art form, Chinese paper-cutting plays a significant role, which is a representative of hollowed art within Intangible Cultural Heritage (ICH)~\cite{Zhang:2021:dot, Wang:2021:dap, ich2009unesco}. It has acquired rich cultural connotations and diverse functions through historical evolution and regional changes~\cite{Cao:2023:the, Ma:2010:sof, Cui:2016:sot}. 
Nowadays, paper-cutting applications have broadened to include various contemporary themes \cite{Cao:2023:the, Liu:2009:rai, Ma:2010:sof}. Despite these advancements and growing influence, it still faces challenges in the vitality of creation and inheritance.
Modern life influenced traditional elements in paper-cutting, leading to their marginalization when they lose relevance to contemporary social contexts~\cite{Wang:2021:dap, Zhang:2018:sos}. Concurrently, the growing trend towards realism in paper-cutting further contributed to the decline of historical and cultural elements~\cite{Cui:2016:sot, Zhang:2018:sos}. Furthermore, paper-cutting with scarce documentation~\cite{Ma:2010:sof, Hu:2017:tda, Huang:2012:ICHprotecting} limits the resources available to the public and even professional practitioners. It also hinders the understanding of the rich cultural knowledge inherent in paper-cutting. These issues exacerbate the homogenization of form and content~\cite{Zhang:2018:sos}, which affects its creativity and inheritance.

With advances in digital technology, computer graphics (CG) show the potential to help with paper-cutting design. Several studies have utilized algorithms for the pattern generation~\cite{Li:2020:aug, Zhang:2006:cpc}, transfer~\cite{Liu:2009:rai, Meng:2010:apc}, and interactive combination for paper-cutting design~\cite{Liu:2018:pdf, Zhang:2005:cag}. However, most approaches depend on stylized pattern generation, which is typically irreversible, and the variation in patterns is often inadequate due to reliance on random seeds~\cite{Liu:2009:rai, Li:2020:aug, Meng:2010:apc}. Although interactive systems~\cite{Liu:2018:pdf, Hu:2017:tda, Zhang:2005:cag} offer some flexibility to enhance content diversity, they are frequently confined to predefined forms and styles, constraining the creative and cultural expression in paper-cutting.

Generative AI (GenAI) for text and images (e.g., ChatGPT\footnote{\url{https://openai.com/chatgpt}}~\cite{chatgpt} and Stable Diffusion\footnote{\url{https://github.com/CompVis/stable-diffusion}}~\cite{Rombach:2022:stablediffusion}), with the comprehensive knowledge base and efficient production capabilities, becomes increasingly integral to the design process~\cite{Heyrani:2021:creativegan, Centinic:2022:aiartreview, Muller:2022:genaichi} to support creative work, such as writing~\cite{Reza:2024:abscribe, Yuan:2022:wordcraft}, painting~\cite{Xu:2023:magicalbrush, Fan:2024:contextcam, Xu:2024:fuzzypainting}, and handicraft design~\cite{Yao:2024:shadowmaker}. 
While GenAI provides promising solutions for artistic creation and design, its application to traditional art, such as paper-cutting, remains largely unexplored and presents specific challenges. GenAI is primarily influenced by source data and input, which can lead to homogeneous and biased outputs. This limitation restricts divergent ideation and fails to adequately manage control within the design process~\cite{Brown:2020:fewshotlearners, Amderson:2024:homogenization, Hou:2024:c2ideas, Xu:2023:magicalbrush, Wang:2024:roomdreaming}. Additionally, GenAI lacks domain-specific knowledge~\cite{Hou:2024:c2ideas, Lu:2023:humanstillwin} and understanding of the cultural context essential for paper-cutting, such as symbolic meanings of a theme across cultures, selecting and combining appropriate patterns to match design intents. These are critical factors of paper-cutting design that existing research has not fully studied.

This research aims to explore the design workflow and the challenges involved in aligning visual form and connotation in paper-cutting design with GenAI assistance, and subsequently identify a design space for paper-cutting. To achieve this, we conducted a two-step formative study involving iterative collaboration with experts. This study included semi-structured interviews with seven participants from different backgrounds to understand their design challenges. It also included a content analysis of 140 paper-cuttings to investigate the pattern and ideation factors within paper-cutting. 

The formative study outlined the workflow (ideation and composition in \autoref{fig:teaser}) and identified challenges inherent in paper-cutting design. Novices face difficulties due to insufficient foundational knowledge, which hinders effective ideation. While experts are proficient in defining processes and ideas to meet requirements, they often rely on experience, leading to fixation. GenAI-aided design struggles with understanding the complex cultural knowledge inherent in paper-cutting. As a result, the outputs are often restricted to fixed or even erroneous output that fails to align with design requirements and cultural contexts. Users are overwhelmed by the inconsistent quality of results in exploration. Additionally, low user engagement with direct text-to-image generation restricts modification and editing, limiting flexibility in creating paper-cutting. 
The content analysis identified four core factors in paper-cutting ideation: \textbf{Function, Subject Matter, Style,} and \textbf{Method of Expression}, and a key element of paper-cutting: \textbf{Pattern}, in various forms, are closely linked to these factors, serving as basic glyph units, primary or decorative components in a paper-cutting.

Building on the identified challenges and the derived design space, we introduce three design goals for a GenAI-aided system aimed at harmonizing form and connotation in paper-cutting design: (1) facilitate users in formulating requirements based on four factors of design, (2) explore the recommended diverse and related contents for translating the idea into a visual representation, (3) enable user's interactive content combination and modification on generated results to encourage their creativity.
Aligned with these design goals, we propose a GenAI-aided paper-cutting design pipeline (\autoref{figure2}) and develop a prototype system, \textbf{HarmonyCut} (\autoref{figure3}), which facilitates user ideation and suggests related content with its corresponding meaning and cultural background, which serves as a reference to support the users in exploring and composing a harmonious visual paper-cutting design.

To evaluate our workflow and system, we conducted a within-subjects user study with 16 participants to compare HarmonyCut with the baseline tool, supported by ChatGPT and DALL-E-3\footnote{\url{https://openai.com/index/dall-e-3/}}, and an expert interview with 3 professional paper-cutting inheritors. The results demonstrated that factor-oriented guidance and reference-based exploration, when integrated with GenAI and domain knowledge, can effectively facilitate ideation and composition in paper-cutting design.





This research thus contributes to the following:
\begin{itemize}
    \item A \textbf{formative study} involving seven participants that identifies the design workflow and challenges inherent in paper-cutting design.
    \item A \textbf{content analysis} of 140 paper-cuttings, which uncovers the core factors and various patterns that define the design space in paper-cutting. 
    \item A \textbf{prototype system}, HarmonyCut, which leverages GenAI to support reference recommendation and exploration for conceptual ideation and visual composition in paper-cutting design.
    \item A \textbf{user study} (N=16) with the \textbf{expert interview} (N=3), exploring how HarmonyCut supports each step of the paper-cutting design process with GenAI assistance.
\end{itemize}

\section{Background and Related Work}
This work aims to assist users in designing paper-cuttings by leveraging GenAI for reference exploration to enhance creativity. In this section, we review previous literature on the following topics: (1) the background of paper-cutting, its current challenges in creation and inheritance, and works that utilize computer graphics and computation-focused methods to aid in paper-cutting design and creation; (2) previous GenAI-aided system in facilitating creativity, particularly in traditional arts; and (3) the use of reference exploration in fostering creative design.

\subsection{Chinese Paper-cutting}

\subsubsection{Background of Chinese Paper-cutting}

Chinese paper-cutting, a significant element of the world's ICH~\cite{ich2009unesco}, is known for its affordability and diverse forms that carry profound cultural connotations~\cite{Qiao:2011:liveICH, Bai:2003:stylistic, Hu:2017:tda}. Physically, paper-cuttings are basically two-dimensional figures hollowed on single-color paper. Culturally, they originate from people's labor practices, serving as an artistic representation of daily life and aspirations~\cite{Cui:2016:sot, Wang:2021:dap, Wang:2021:folk}. These works embody human aesthetic concepts and ideals, incorporating religious beliefs, moral values, and cultural elements~\cite{Cao:2023:the}.
Over centuries, paper-cutting has surpassed regional and ethnic boundaries, integrating into local customs while preserving its form with connotation~\cite{Zhang:2018:sos, Ma:2010:sof}.


With the acceleration of modernization and urbanization, paper-cutting faces significant challenges in themes, content, and talent cultivation, affecting its innovation and inheritance~\cite{Zhang:2018:sos, Li:2023:digitalpapercut, Wang:2021:dap}. The content of paper-cuttings increasingly reflects modern life and realism~\cite{Zhang:2018:sos, Cao:2023:the}, leading to the gradual disappearance of traditional themes and their symbolic meanings. The economic returns from modern paper-cutting are relatively low, making it difficult to sustain as a profession, resulting in fewer individuals pursuing it. Additionally, as a form of folk art~\cite{Wang:2021:folk}, paper-cutting primarily relies on oral and practical teaching and lacks systematic documentation, complicating its transmission~\cite{Wang:2018:transinheritance} and making it inaccessible for both the public and practitioners for reference. Consequently, this intensifies the homogenization and stylization of paper-cutting themes with diminished meaning, thereby inhibiting innovation in design and creation~\cite{Zhang:2018:sos}.

\subsubsection{Paper-cutting Design with Digital Technology }
Advancements in digital technology have facilitated the use of CG and computer vision (CV) in paper-cutting design and creation. Several studies have concentrated on analyzing geometric features, such as symmetrical and geometric motifs, to build pattern libraries by extracting and classifying patterns~\cite{Zhang:2005:cag, Zhang:2006:cpc, Zhang:2009:cutout, Shui:2008:edgepaper, Liu:2018:pdf, Li:2020:aug}. Some have even utilized algorithms to enhance various pattern forms and styles~\cite{Liu:2009:rai}. Nonetheless, these approaches are constrained in generating designs with complex shapes and diverse meanings, resulting in a limited range of categories and styles.
Image-based methods employ multi-layer image processing algorithms to create connected paper-cutting images directly~\cite{Xu:2007:computer, Xu:2008:artistic, Hu:2017:tda}. Despite their utility, these methods often struggle to produce meaningful content that matches design needs and specific shapes, sometimes compromising aesthetics to maintain connectivity.
Some works have applied CV techniques, such as utilizing the convolutional neural network for pattern recognition~\cite{Liu:2020:intcut} and transfer learning to transform portraits into paper-cutting styles~\cite{Meng:2010:apc}. These methods face difficulties in processing abstract paper cuttings with unique styles and deep meanings. In fact, many related fields, such as graphic design, have already started integrating GenAI into their systems. To the best of our knowledge, no prior work has investigated the use of GenAI to support paper-cutting design. By examining how GenAI supports creative design in these areas, we aim to leverage this approach and, through the integration of the paper-cutting knowledge we have collected, realize GenAI-aided paper-cutting design.

\subsection{Generative AI for Creativity}
GenAI, leveraging advancements in language models~\cite{Radford:2018:gpt2, Radford:2019:gpt2}, image generation models~\cite{Kingma:2022:vae, Goodfellow:2020:gan}, and multi-modal models~\cite{Radford:2021:clip, Li:2022:blip, Rombach:2022:stablediffusion}, employs its extensive training resources and capabilities in multi-modal information alignment to effectively understand and generate diverse content, thereby facilitating user creativity. In the realm of creative design, proposed language models~\cite{Swanson:2021:story}  are utilized to provide users with abundant ideas and text suggestions during the writing process. Meanwhile, recent text-image models~\cite{Chung:2024:styleid, Zhang:2023:inst, Chen:2024:democaricature, Zhang:2023:prospect}, with their comprehension of style, position, and color, enable the translation of user ideas into images across various scenarios. This synergy of language and image models enhances the creative process by offering integrated support for both ideation and visual generation.






\subsubsection{Creative Support Tool with Generative AI}
To integrate GenAI into specific areas within a controllable process that includes human involvement, numerous frameworks and interfaces have been proposed to support GenAI-aided creativity.
In the realm of graphic design, which is closely related to paper-cutting design, there is a growing trend of integrating GenAI with interfaces to support more specialized and diverse creative tasks. For ideation, tools like GANCollage~\cite{Wan:2023:gancollage} and CreativeConnect~\cite{Choi:2024:creativeconnect} utilize GenAI-driven digital mood board to facilitate brainstorming and recombine references for the creative idea.
Regarding visual representation, some works use GenAI, such as PromptCharm~\cite{Wang:2024:promptcharm} and ContextCam~\cite{Fan:2024:contextcam} to enable multi-modal models with refined prompt and context awareness to generate artistic images that meet various needs. 
Additionally, in specific areas such as communication, emotional affect, and typography, many projects utilize GenAI to suggest ideas and transform them into final visual designs. For instance, the work of Chen et al.~\cite{Chen:2024:dataanalogy} and the Opal~\cite{Liu:2022:opal} both leverage GenAI to generate and apply suggestions for effective illustrations creation. Wan et al.~\cite{Wan:2024:metamorpheus} use GenAI to develop metaphorical suggestions, supporting visual storytelling of emotional experiences. TypeDance~\cite{Xiao:2024:typedance} empowers users to design semantic typographic logos from customized images through a structured workflow.
In traditional art creation, which is more closely related to paper-cutting, Magical Brush~\cite{Xu:2023:magicalbrush} generates symbolic cultural content with GenAI and offers a predefined material library to assist novices in creating modern Chinese paintings. ShadowMaker~\cite{Yao:2024:shadowmaker} enables designers to generate shadow puppet images through style transfer of sketches, ultimately combining them to create shadow puppet animations.

\revisedtext{Although GenAI demonstrates potential in supporting creative tasks, it also faces limitations. A key issue is the end-to-end nature, which limits user involvement in the creative process. Creative tasks often rely on iterative refinement, but GenAI lacks the flexibility to provide control over iterations or ensure alignment with user expectations~\cite{Hou:2024:c2ideas, Brown:2020:fewshotlearners, Duvsek:2020:end-to-end}. Another major limitation arises from the model's dependence on data. In tasks requiring domain-specific knowledge, GenAI frequently underperforms~\cite{Hou:2024:c2ideas, Cui:2024:chatlaw, Wang:2023:methodsknowledge}. Moreover, biases in training data often result in homogenized outputs~\cite{Amderson:2024:homogenization}, reducing diversity and originality while limiting users to explore and innovate. These biases may even lead to issues about copyright or ethical concerns, further complicating the creative process~\cite{Zhou:2024:biasgenerativeai, Samuelson:2023:copyright}. To address these challenges, we restrict GenAI's role to providing partial reference support during the design process and incorporate predefined elements to reduce model unpredictability. Furthermore, as GenAI has yet to fully explore the domain-specific knowledge and design space of paper-cutting, we focus on advancing this area and aim to guide users through reference-based exploration while improving controllability and rationality~\cite{Hou:2024:c2ideas}.}



\subsubsection{Reference Exploration in Design}


An effective design process requires users to explore multiple alternatives~\cite{Jansson:1991:designfixation, Goldschmidt:2011:avoidingfixation}, and GenAI is employed for creative support due to its capability to offer numerous high-quality variations. However, some challenges persist: GenAI-generated content is often end-to-end complete, potentially causing users to become overly reliant on the model, merely selecting the model's suggestions instead of fostering genuine creativity~\cite{Tohidi:designreight}. Furthermore, although a vast amount of content can be produced based on user requirements, without proper guidance, users may feel overwhelmed by the excess of options~\cite{Suh:2024:luminate}, making it challenging to decide which elements to incorporate into their designs. A key step in the design process is drawing inspiration from references~\cite{Shneiderman:2000:creating, Eckert:2000:sources}, which assists users in navigating relevant design spaces. This strategy ensures that the design process is not solely dependent on fully developed generated content, allowing for the creation of designs based on rough yet innovative ideas~\cite{Tohidi:designreight, Choi:2024:creativeconnect}, and also mitigates the risk of overwhelming users with too many creative options. 
For the aforementioned issues, previous work has demonstrated that mood boards~\cite{Cassidy:2008:moodboards} can help users explore and be guided by reference and inspiration~\cite{Eckert:2000:sources, Garner:2001:problem}. Moreover, an increasing number of GenAI-aided systems support design by utilizing interactive mood boards~\cite{Choi:2024:creativeconnect, Wan:2023:gancollage, Peng:2024:designprompt}. Inspired by previous work, our research facilitates reference-guided exploration in two primary steps of the paper-cutting workflow and utilizes a mood board to support users in the selection, arrangement, and cutout of these references for paper-cutting design.

\section{Formative Study}\label{sec:formative}
We conducted a formative study to understand the workflow of paper-cutting design and users' challenges during the design.


\subsection{Participants}
As a work targeted to public users, we initially sought to examine the paper-cutting design process and the challenges encountered without GenAI. Subsequently, we aimed to assess the role and limitations of integrating GenAI into paper-cutting design. To achieve this, we recruited participants with varying levels of expertise in both paper-cutting and GenAI, as experts are expected to have insights into design workflows, and each participant group faces different challenges in paper-cutting design. \rrtext{Participants were categorized into four levels of paper-cutting expertise, with further clarification regarding the criteria for paper-cutting expertise provided in~\autoref{expert clarification}:}
(1) \textbf{Masters}: who have over 20 years of professional experience in paper-cutting creation are officially recognized as ICH inheritors; (2) \textbf{Practitioners}: who have 10-20 years of experience in paper-cutting-related work; \revisedtext{(3) \textbf{Amateurs}: who have 1-3 years of experience in creating paper-cuttings without systematic training;} (4) \textbf{Novices}: who never engaged in paper-cutting design or creation. Additionally, we defined three levels of GenAI expertise: (1) \textbf{Professionals}: researchers in the multi-modal machine learning field; 
\revisedtext{(2) \textbf{Knowledgeable Users}: who regularly integrate GenAI into their professional, educational, or personal activities and have experience with basic prompt engineering. (3) \textbf{Novices}: who have minimal exposure to GenAI, may have heard of it without engaging in its use, or are entirely unacquainted with it.}

\revisedtext{Seven participants (3 females and 4 males; age M=34.43, SD=12.79) were recruited for the semi-structured interviews through online postings on various social media platforms (e.g., Douyin\footnote{\url{https://www.douyin.com/}} and Bilibili\footnote{\url{https://www.bilibili.com/}}), and the detailed information of the participants is shown in \autoref{table:formative participants}.} Among them were three paper-cutting masters (P1, P4, P5), two practitioners (P2, P3), one amateur (P7), and one novice (P6). Regarding GenAI expertise, the group included one GenAI professional (P6), two knowledgeable users (P2, P7), and four novices (P1, P3, P4, P5). Each participant is compensated with 100 CNY (approximately 14 USD).

\subsection{Procedure}
The semi-structured interview included two parts: (1) individual design and (2) GenAI-aided design.
\revisedtext{At the beginning, we provided separate 10-minute background introductions for novices in the fields of paper-cutting and GenAI, amounting to a total of 20 minutes.}
In the initial part of the study, each participant was tasked with providing a design concept description for paper-cuttings and giving a sketch of paper-cutting, including the areas to cut out, selecting randomly from two main themes:  ''\textit{Dragon Boat Festival}'' and ''\textit{Wedding,}'' for 30 minutes. We then interviewed them to explore participants' personal understanding and perspectives on the paper-cutting design process and examined the key aspects of the design process, including steps, core factors, and challenges faced.
Then, based on the aforementioned themes, participants were asked to engage with a Large Language Model (LLM) to assist in the design process and use a Text-to-Image model to generate paper-cutting image in 10 minutes. After that, we collected feedback from participants on GenAI-aided paper-cutting design, including the shortcomings of the results, challenges in the design process, and their expectations and suggestions for the GenAI-aided system.




\subsection{Findings}

Through the semi-structured interviews and literature review, we identified the workflow of paper-cutting design: ideation and composition. We found that ideation in paper-cutting design presents challenges to all participants, albeit to varying degrees. It is tedious to both participants and GenAI in the composition phase. In GenAI-aided designs, the recommended and generated content is often undesirable, leading to an uncontrollable overall process that cannot modify the output.

\subsubsection{Workflow of Paper-cutting Design}
\revisedtext{Drawing on feedback from interviews, the suggested process of paper-cutting creation, especially in paper-cutting education~\cite{Lin:1974:howtopapercutting, Li:1998:monopapercutting, Li:2011:PatternandDesign, Zhang:1982:discusspapaercut}, and the design steps from Hubka et al.~\cite{Hubka:1992:engineeringdesign}, paper-cutting design primarily involves two main steps, as shown in \autoref{figure1}~(A).} The first step is transforming intents into a conceptual idea~\cite{Choi:2024:creativeconnect}.
The second step is translating the conceptual idea into visual form (a design blueprint before using scissors for cutting).
\begin{itemize}
\item \textbf{Ideation.} The first step involves transforming intents with a theme into a conceptual idea~\cite{Li:1998:monopapercutting}. 
\rrtext{Based on the experts' feedback, several preliminary dimensions were mentioned, including \textbf{function and style}. These dimensions suggest that ideation should be approached from multiple dimensions (4 factors as detailed in \autoref{sec:4_2}) to determine the core components of the design.} This idea will later be translated into a visual design in the next step.
\item \textbf{Composition.} During the composition step, designers (1) select the shapes of the elements based on the idea, (2) arrange and combine the selected contours, and (3) decide on cut-out regions (unit patterns) for future creation~\cite{Lin:1974:howtopapercutting, Li:2011:PatternandDesign}.
\end{itemize}

\subsubsection{Challenges in Paper-cutting Design}
\revisedtext{Based on the observation in the formative study and literature review related to design and paper-cutting, we refined and summarized 5 challenges in paper-cutting design with GenAI assistance.}
\begin{itemize}
\item[\textbf{C1.}] \textbf{Challenges in Translating Intent to Ideas.}
Novices and amateurs (P6 and P7) spent significant time on the first part of the study but could only provide two vague descriptions of their design concepts for each theme.
\revisedtext{This difficulty arises not only from their limited knowledge of paper-cutting, including familiarity with the design workflow and the essential elements needed to address the theme but also from the absence of a cognitive approach (i.e., ``mapping thinking'') described by Li~\cite{Li:1998:monopapercutting}. This approach enables them to creatively link their knowledge to the attributes of natural objects, assign meaningful concepts, and employ structural methods to effectively express their paper-cutting design ideas. All these competencies are essential to learn and apply for paper-cutting creation~\cite{Lin:1974:howtopapercutting, Zhang:2021:papercuttingteaching}.}
As noted by P7, ``\textit{It is easy to cut a paper-cutting based on a sketched outline, but besides the content, I am unsure of which dimensions need consideration in the design process.}'' \revisedtext{Besides, P6 stated, ``\textit{I don't know what content in paper-cutting can appropriately map to those creation intents.}''} Consequently, it is challenging to establish a clear direction for their ideas.

\item[\textbf{C2.}] \textbf{Lack of Creativity and Multiple Variations in Ideation.}
\revisedtext{Novices and amateurs (P6 and P7), who possessed limited knowledge of paper-cutting, struggled to select elements aligned with their themes, and their proposed ideas appeared repetitive. These findings are consistent with prior research~\cite{Ericsson:1994:knowledgecreative, Gardner:1993:knowledgecreative}, which highlight that long-term immersion in a discipline is essential for developing creative ability. Furthermore, knowledge is identified as a critical foundation for fostering innovative ideas~\cite{Weisberg:1999:creativityandknowledge}, supporting the finding that a lack of systematic knowledge hinders the creative potential of novices and amateurs in paper-cutting design.}
Conversely, experts and practitioners, although proficient in the creative process and capable of rapid ideation, tend to rely heavily on their accumulated experience and local cultural influences for themes and content. This dependence can lead to fixation on a single idea. As P4 said, ``\textit{Paper-cutting is highly regional, with the meaning of specific elements differing significantly even within the same province. Although there is diversity in paper-cutting, I am only familiar with the themes and elements from my region, resulting in more fixed forms and content across many themes.}''

\item[\textbf{C3.}] \textbf{Challenges in Converting Ideas into Concrete Visual Representations.}
Novices have limited drawing skills, making translating content in their ideas directly into visual forms challenging. Additionally, composition requires considering not only the spatial arrangement and structure of elements but also deciding which areas should be cut-out (pattern) and in what shapes during creation. It is labor-intensive for both novices and experts. As mentioned by P1, ``\textit{The arrangement of specific content and the shapes created through cut-outs (pattern) best reflect personal style. However, translating an idea into a visual expression is still laborious.}''

\item[\textbf{C4.}] \textbf{Challenges in Exploring Suitable and Rational GenAI Results.} 
For the given themes, the model struggles to grasp the user's unique design idea, often providing overly broad suggestions, which cannot assist the user in avoiding fixation even increase the load to user in exploration. Additionally, the finally generated paper-cutting images often do not match the text description, especially in the spatial arrangement of content. Moreover, many parts of the model-generated paper-cuttings are irrational or irrelevant, such as generating some random clusters of stripes as patterns in paper-cutting.

\item[\textbf{C5.}] \textbf{Challenges in Controlling and Editing GenAI Results.} 
Regarding the above issue with GenAI, it is difficult to directly adjust errors in the generated results. Participants can only try to improve the output by revising the input descriptions. However, because the model struggles to understand the knowledge of contents with nuanced meaning and composition, iterative changes to the input often yield minimal improvement.

\end{itemize}



\begin{table*}[!htbp]
\caption{Definition and examples of factors and types in paper-cutting ideation derived from content analysis.}
  \Description{This table demonstrates the definition and examples of factors and types in paper-cutting ideation derived from content analysis.}
  \label{table1}
\Large
\renewcommand\arraystretch{1.7}
\resizebox{\textwidth}{!}{
\begin{tabular}{l|l|l|p{8.cm}|l}
\hline
\multicolumn{1}{c|}{\textbf{Category}}                                          & \multicolumn{1}{c|}{\textbf{Subcategories}} & \multicolumn{1}{c|}{\textbf{Type}} & \multicolumn{1}{c|}{\textbf{Definition}}                                          & \multicolumn{1}{c}{\textbf{Examples (Figures)}} \\ \hline
\multirow{7}{*}{Function}                                                       & \multirow{4}{*}{Spiritual Function}         & Witchcraft Belief                     & Serve as a symbol in witchcraft activities, embodying related beliefs and rituals    &  Wizard Exorcising Demons~(\autoref{a1fig1}(a))  \\ \cline{3-5} 
                                                                                &                                             & Indigenous Belief                  & Reflect unique local belief systems as a form of cultural expression              & Herding Ducks in Watertown~(\autoref{a1fig1}(b))  \\ \cline{3-5} 
                                                                                &                                             & Religious Belief                    & Act as a symbol in religious ceremonies or doctrines, conveying religious content & Guanyin Sitting on a Lotus~(\autoref{a1fig1}(c))  \\ \cline{3-5} 
                                                                                &                                             & Cultural Dissemination                & Serve as an medium to disseminate culture and historical information              & Happy Asian Games~(\autoref{a1fig1}(d))  \\ \cline{2-5} 
                                                                                & \multirow{3}{*}{Practical Function}         & Interpersonal Communication        & Act as a medium in social etiquette settings in interpersonal communication       & Mandarin Ducks in Water~(\autoref{a1fig1}(e))  \\ \cline{3-5} 
                                                                                &                                             & Festive Atmosphere Evoking        & Enhance the atmosphere and cultural features in holiday or seasonal celebrations  & Solar Term: Grain Full~(\autoref{a1fig1}(f))  \\ \cline{3-5} 
                                                                                &                                             & Daily Decoration                   & Serve as decorative items in daily life                                           & Butterfly Window Decoration~(\autoref{a1fig1}(g))  \\ \hline
\multirow{6}{*}{Subject Matter}                                                        & \multirow{5}{*}{Traditional Subject Matter}        & Primitive Paper-cutting            & Present paper-cutting by initial form in history                               & Circular Floral Paper-cutting~(\autoref{a1fig1}(h))  \\ \cline{3-5} 
                                                                                &                                             & Flora and Fauna                    & Present paper-cutting by animals and plants                                       & The World Welcomes Spring~(\autoref{a1fig1}(i))  \\ \cline{3-5} 
                                                                                &                                             & Landscape                         & Present paper-cutting about natural and cultural landscapes                       & Lijiang Ancient Town~(\autoref{a1fig1}(j))  \\ \cline{3-5} 
                                                                                &                                             & Historical Figure and Story                    & Present paper-cutting centered around stories of characters                       & Jiang Ziya Fishing~(\autoref{a1fig2}(a))  \\ \cline{3-5} 
                                                                                &                                             & Folk Life                & Present paper-cutting about traditional customs and culture                      & The Mouse's Wedding~(\autoref{a1fig2}(b))  \\ \cline{2-5} 
                                                                                & Innovative Subject Matter                         & Contemporary Subject               & Paper-cutting integrated with modern subjects                                     & Genshin Impact, Klee~(\autoref{a1fig2}(c))  \\ \hline
\multirow{2}{*}{Style}                                                          & Abstract Style                              & -                                  & Express ideas with non-representational forms by paper-cutting                   & Frog,~\autoref{a1fig2}(d)  \\ \cline{2-5} 
                                                                                & Realistic Style                             & -                                  & Replicate real objects and scenes by paper-cutting                                & Lujiazui, Shanghai~(\autoref{a1fig2}(e))  \\ \hline
\multirow{3}{*}{\begin{tabular}[c]{@{}l@{}}Method of\\ Expression\end{tabular}} & Metaphor                                    & -                                  & Use similar or related content to indirectly express meaning or emotion         & Harmonious of Ethnicities~(\autoref{a1fig2}(f))  \\ \cline{2-5} 
                                                                                & Symbolism                                   & -                                  & Use specific content to represent theme                                            & Enduring Lineage~(\autoref{a1fig2}(g))  \\ \cline{2-5} 
                                                                                & Homophony                                   & -                                  & Use similarity of pronunciation to embed positive wishes into specific objects    & Happiness Arrives~(\autoref{a1fig2}(h))  \\ \hline
\end{tabular}
}
\end{table*}

\section{Content Analysis of Paper-cuttings and Patterns}\label{sec:content}
Based on the formative study, we discovered that there are some key aspects to consider in the design workflow. To identify them, we first collected a paper-cutting dataset. We conducted two content analyses: one to develop an ideation factors taxonomy and another for a pattern taxonomy (taxonomy for the key element in ideation and composition). \rrtext{Both analyses were carried out under the guidance of expert review, involving five experts (P1–P5) who had also participated in the formative study. The experts were involved in two stages: (1) During the coding process, ambiguities in the instance-level annotation of paper-cuttings and patterns were resolved through a single expert-guided discussion conducted via the WeChat group, with the final annotation determined based on the majority vote of all experts; (2) Additionally, after each version of the codebook was drafted by the authors, the experts participated in discussions conducted via online conferences. In each round, the experts collaboratively reviewed and evaluated the current version of the codebook, offering refined or expanded suggestions, which were systematically discussed and consolidated to resolve ambiguities and ensure alignment of perspectives. This iterative process continued until all experts reached a consensus on the coding results, ensuring that the final codebook was validated.}

\begin{table*}[!htbp]
\caption{Definition and examples of pattern categories and subcategories derived from content analysis.}
  \Description{This table demonstrates the definition and examples of pattern categories and subcategories derived from content analysis.}
  \label{table2}
\resizebox{\textwidth}{!}{
\begin{tabular}{c|c|p{8.1cm}|cc}
\hline
\textbf{Category}                            & \textbf{Subcategory}     & \makecell*[c]{\textbf{Definition}} & \multicolumn{2}{c}{\textbf{Examples}} \\ \hline
\multirow{3}{*}{Unit Pattern~\cite{Hu:2021:Traditionalpattern, Zhuge:1998:patterndictionary}}       & Geometric Unit Pattern       & \makecell*[c]{Independent units based on abstract geometric\\ glyphs that can be integrated with other patterns \\in paper-cuttings~\cite{Tian:2003:Historypattern, Liu:2009:rai, Liu:2020:intcut, Li:2020:aug}}     &               \raisebox{-.39\dimexpr\totalheight-\ht\strutbox}{\includegraphics[scale=0.69]{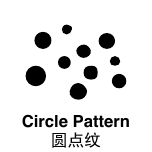}} &  \raisebox{-.4\dimexpr\totalheight-\ht\strutbox}{\includegraphics[scale=0.69]{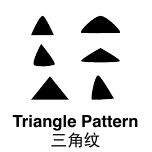}}           \\ \cline{2-5} 
                                    & Semantic Unit Pattern          & \makecell*[c]{Independent units based on specific semantic glyphs \\ that can be integrated with other patterns in paper-cuttings}     &             \raisebox{-.4\dimexpr\totalheight-\ht\strutbox}{\includegraphics[scale=0.69]{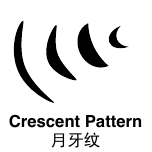}} &  \raisebox{-.43\dimexpr\totalheight-\ht\strutbox}{\includegraphics[scale=0.69]{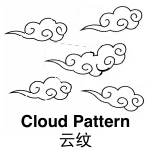}}           \\ \cline{2-5} 
                                    & Sawtooth Pattern   & \makecell*[c]{Specific independent units resembling sawtooth, \\formed by replicating various glyphs along\\ a trajectory for depicting contour and light gradient \\in paper-cuttings~\cite{Zhang:2018:sos, Liu:2020:intcut, Zhang:2006:cpc, Liu:2009:rai}}       &            \raisebox{-.51\dimexpr\totalheight-\ht\strutbox}{\includegraphics[scale=0.72]{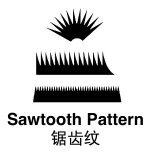}} &  \raisebox{-.51\dimexpr\totalheight-\ht\strutbox}{\includegraphics[scale=0.72]{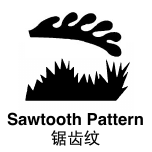}}             \\ \hline
\multirow{2}{*}{Composite Pattern} & Primary Composite Pattern        & \makecell*[c]{Composite patterns made up of multiple unit patterns \\ that serve as the primary objects for \\ expressing themes in paper-cuttings}        &              \raisebox{-.5\dimexpr\totalheight-\ht\strutbox}{\includegraphics[scale=0.66]{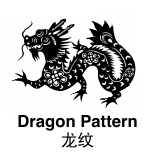}} &  \raisebox{-.5\dimexpr\totalheight-\ht\strutbox}{\includegraphics[scale=0.66]{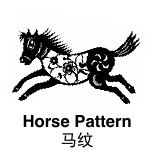}}             \\ \cline{2-5} 
                                    & Decorative Composite Pattern &  \makecell*[c]{Composite patterns made up of multiple \\unit patterns that decorate primary objects \\in paper-cuttings~\cite{Zhuge:1998:patterndictionary, Hu:2021:Traditionalpattern, Zhang:2005:cag}}       &             \raisebox{-.51\dimexpr\totalheight-\ht\strutbox}{\includegraphics[scale=0.66]{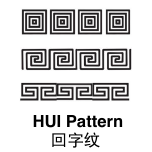}} &  \raisebox{-.43\dimexpr\totalheight-\ht\strutbox}{\includegraphics[scale=0.66]{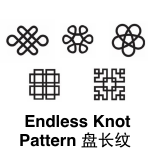}}            \\ \hline
\end{tabular}
}
\end{table*}
\subsection{Data Collection}\label{sec:data}
\revisedtext{We collected over 17,000 paper-cuttings through data crawling from the Chinese Paper Cutting Digital Space\footnote{\url{https://www.papercutspace.cn/}}, which organized and merged paper-cuttings into seven distinct categories based on the human geography regions in China~\cite{Fang:2017:chinahumangeo} (i.e., Central China, East China, North China, Northeast, Northwest, South China, and Southwest).} We then selected 1,521 paper-cuttings with titles from the source to aid in identifying the content, as some abstract works are challenging to interpret and annotate in content analysis. We further filtered out low-definition images and multi-color paper-cuttings, as our focus is on traditional monochromatic Chinese paper-cutting. This process resulted in a collection of 701 paper-cutting images, which serve as the sampled source for content analysis in~\autoref{sec:4_2} and \autoref{sec:4_3}. Additionally, these images function as the retrieval dataset discussed in~\autoref{sec:harmonycut}. 
\revisedtext{For content analysis and fine-tuning models, we sampled 20\% images (140/701) based on the distribution of the seven regions to ensure alignment with the regional characteristics of paper-cutting, which is shown in~\autoref{figure:sample filter}.}

\subsection{Core Factors of Paper-cutting Design Ideation} \label{sec:4_2}
\revisedtext{We first collected all dimensions (i.e., function and style) considered by experts during the ideation phase of the formative study based on their feedback (\autoref{sec:formative}). To identify the ideation factors and the various types of content within each factor, the first author, in collaboration with five experts, conducted a content analysis using 140 selected paper-cuttings~(\autoref{sec:data}).
The first author randomly selected 70 (50\%) paper-cuttings to create taxonomy by open-coding. To ensure the quality of the taxonomy, two criteria were followed during the coding process~\cite{Nickerson:2013:Taxonomymethod}: all elements in the taxonomy should be comprehensive, covering every aspect, and the types within each factor should be mutually exclusive.}
In the beginning, the coder annotated detailed content across all dimensions observed or clarified through experts when content was unclear or ambiguous, resulting in the initial codebook. 
\rrtext{For instance, Northeast shamanistic paper-cuttings with the regional characteristic, were initially challenging to determine their function. Following the expert-guided discussion and voting, experts agreed their function in witchcraft beliefs takes precedence over indigenous beliefs.
Then, the coder produced the final codebook under expert review through three rounds of discussion and iterations (the detailed information of each round is shown in~\autoref{A:expert discuss ideation}), identifying 18 types of content grouped into 4 factors:} Function, Subject Matter, Style, and Method of Expression. The categorization results, including ideation factors, sub-factors, types, definitions, and example paper-cuttings, are detailed in~\autoref{table1}. The first author then applied the finalized codebook to code the remaining 70 paper-cuttings and validated the taxonomy, as shown in~\autoref{figure:paper-cut validation}.
Examples of each type are shown in~\autoref{A:content examples}. For the 140 paper-cuttings, each is annotated type and has specific explanations for every factor, such as category \textit{Subject Matter}, type \textit{Historical Figure and Story}, detailed information ``\textit{A paper-cutting narrates the historical story of Zhaojun Wang's journey to the Xiongnu for a political marriage.}'' These 140 pieces are used as domain knowledge and ground-truth datasets in~\autoref{sec:harmonycut}.


\subsection{Patterns in Paper-cutting}\label{sec:4_3}
\revisedtext{From a semiotic perspective, Chinese paper-cutting consists of a series of symbolized patterns~\cite{Liang:2011:Symbolpattern}. Saussure's theory~\cite{Saussure:1916:Semiotics} defines a symbol as comprising two parts: the ``signifier'' and the ``signified.'' In the context of paper-cutting, the ``signifier'' refers to the form or structure of the pattern, while the ``signified'' represents the meaning or concept it conveys~\cite{Cao:2009:Semiotics, Liang:2011:Symbolpattern}. 
Therefore, achieving harmony between form and connotation in paper-cutting design requires a comprehensive understanding of patterns that unify meaning and content. To support this, we developed a taxonomy of these patterns using a hybrid thematic analysis approach, also using those 140 selected paper-cuttings~(\autoref{sec:data}).}

\revisedtext{In the deductive coding process, previous research found three important aspects of pattern and paper-cutting namely \textit{Geometric Pattern}~\cite{Tian:2003:Historypattern, Liu:2009:rai, Liu:2020:intcut, Li:2020:aug}, \textit{Decorative Pattern}~\cite{Zhuge:1998:patterndictionary, Hu:2021:Traditionalpattern, Zhang:2005:cag}, and \textit{Sawtooth Pattern}~\cite{Zhang:2018:sos, Liu:2020:intcut, Zhang:2006:cpc, Liu:2009:rai}. \textit{Geometric Pattern} was composed of shapes formed by the ultimate abstraction of points, lines, and planes. \textit{Decorative Pattern} was used to enhance or embellish specific patterns or the overall composition of a paper-cutting. \textit{Sawtooth Pattern} was a distinct and important pattern in paper-cutting, used to convey texture and layering in objects while also serving as decoration for various patterns. However, these three aspects partially overlap in both shape and function. To address this, we adopted the concept of \textit{Unit Pattern}~\cite{Hu:2021:Traditionalpattern, Zhuge:1998:patterndictionary}, representing the fundamental unit in paper-cutting, to ensure the codebook remained mutually exclusive. Thus, \textit{Geometric Pattern} and \textit{Sawtooth Pattern} were identified as sub-categories of \textit{Unit Pattern}, and \textit{Decorative Pattern} was defined as one of \textit{non-Unit Patterns} with decorative functions.}

\revisedtext{In the inductive coding process, OpenCV extracted 63,452 cut-outs (i.e., \textit{Unit Patterns}) from 140 paper-cuttings~(\autoref{sec:data}), and 1,269 cut-outs (2\%) were randomly sampled from them.}
The first author open-coded 635 cut-outs and 70 paper-cuttings, both randomly selected at 50\%, similar to the previous process in~\autoref{sec:4_2}, referring to pattern knowledge~\cite{Huang:2021:chinesepattern, Wang:2009:folkpattern, Zhao:2023:papercutculture, Zhuge:1998:patterndictionary}.
\rrtext{Ambiguities in instance-level annotation were resolved using the method outlined in~\autoref{A:paper-cut coding}. For example, the classification of the plum blossom pattern as either primary or decorative was challenging. After the single expert-guided discussion, all five experts agreed it is primarily decorative and rarely used as a main subject.}
\rrtext{Through two rounds of expert discussions and iterations (the detailed information of each round is shown in~\autoref{A:expert discuss pattern}), a taxonomy of patterns was established~(\autoref{table2}), consisting of two categories:} \textit{Unit Patterns} and \textit{Composite Patterns} (i.e, \textit{non-Unit Patterns}). In addition, two new subcategories were introduced: unit semantic patterns and composite primary patterns. 
\textit{Unit patterns} include 25 different patterns (8 geometric units, 12 semantic units, and 5 sawtooth patterns). \textit{Composite patterns}, which represent the primary content of paper-cuttings, are formed by combining unit patterns and include 42 different patterns (8 decorative composite patterns and 34 primary composite patterns). The first author annotated the remaining 70 paper-cuttings and 634 cut-outs, and validated taxonomy~(\autoref{figure:pattern validation}). 
\rrtext{All the names and examples of the 25 specific \textit{Unit patterns} and 42 specific \textit{Composite patterns} are provided in the supplementary material for detailed reference.}
These labeled patterns serve as both a repository of domain knowledge and ground truth datasets in \autoref{sec:harmonycut}.

\section{Design Goals}
\revisedtext{Based on the findings from the semi-structured interviews and content analysis, we identified three design goals (\autoref{figure1}~(C)) for developing a GenAI-aided system that supports users in designing a paper-cutting where explicit elements align with implicit meanings:}
\begin{itemize}
    \item[\textbf{DG 1.}] \textbf{Factor-oriented Guidance for Ideation.} This goal aims to help users efficiently transform abstract intent into structured descriptions (\textbf{C1}). The system guides ideation by focusing on four key factors, providing detailed content based on the selected factors to suggest related content for creatively shaping the final idea (\textbf{C2}).

    \item[\textbf{DG 2.}] \textbf{Reference-based Exploration for Composition} For each compositional action, the system based on domain knowledge suggests related content as the reference. Users can explore these design references for creative self-selection and arrangement of contours and patterns extracted from reference (\textbf{C3, C4}).

    \item[\textbf{DG 3.}] \textbf{Controllable and Editable Design with Recommendation} The recommendations provided by GenAI are integral to DG 1 and 2. These suggestions can ensure diversity to foster creativity while maintaining rationality and relevance, thereby preventing users from being misled or overwhelmed during exploration (\textbf{C4}). Furthermore, users are able to edit selections of unreasonable or partially reasonable ideas and compositions, allowing them to incorporate desired elements into their designs and complete the entire design process (\textbf{C5}).
    
    
    
\end{itemize}




\begin{figure*}[!htbp]
\centering
\includegraphics[width=0.98\textwidth]{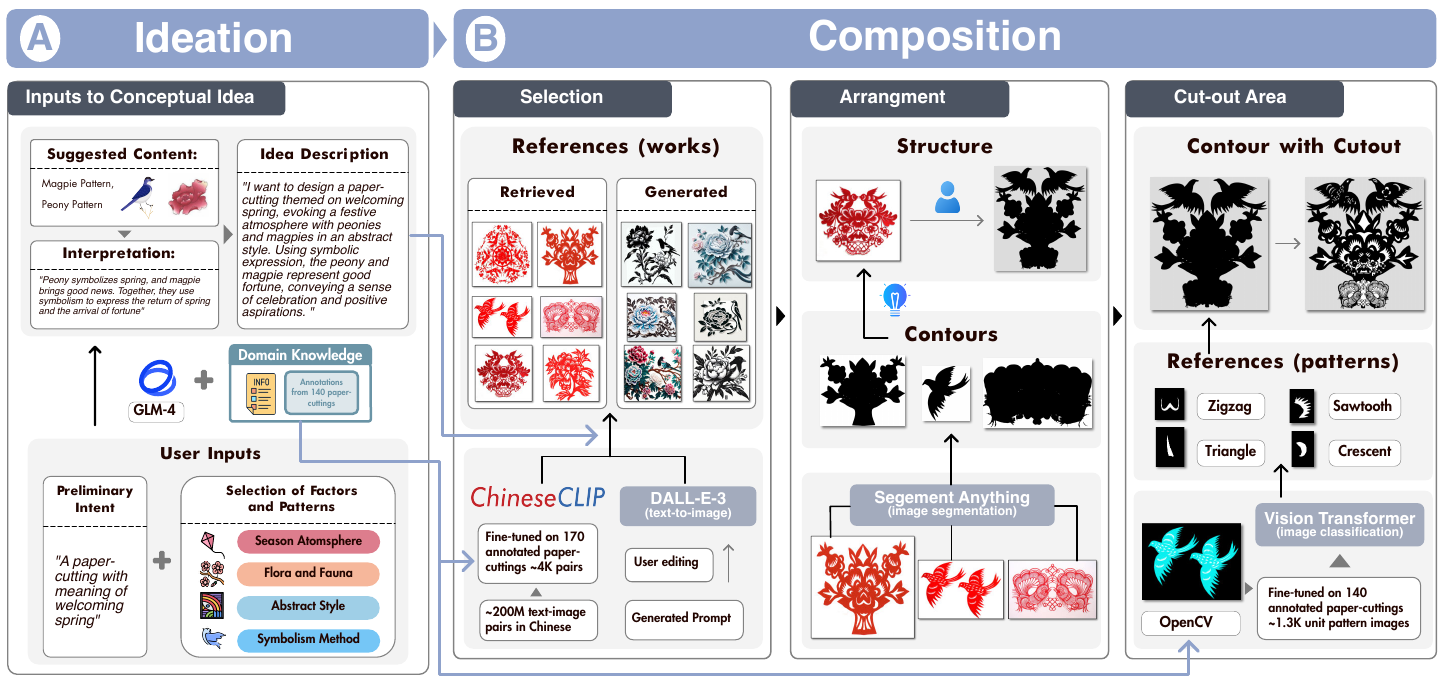}
\caption{\label{figure2}
The pipeline consists of Ideation and Composition components, structured around the summarized workflow and design space. The Ideation component offers knowledge-based guidance, allowing designers to explore and select content that aligns with their intent to form ideas. These ideas are then fed into multi-modal models within the Composition component, which retrieves and generates related content as references. This exploration of references empowers users to arrange reference and plan cut-out areas, facilitating the composition of their own paper-cutting designs.}
\Description{This figure shows the pipeline consists of Ideation and Composition components, structured around the summarized design space and workflow. The Ideation component offers knowledge-based guidance, allowing designers to explore and select content that aligns with their intent to form ideas. These ideas are then fed into multi-modal models within the Composition component, which retrieves and generates related content as references. This exploration of references empowers users to arrange reference and plan cut-out areas, facilitating the composition of their own paper-cutting designs.}
\end{figure*}

\section{HarmonyCut}\label{sec:harmonycut}
\begin{figure*}[!htbp]
\centering
\includegraphics[width=\textwidth]{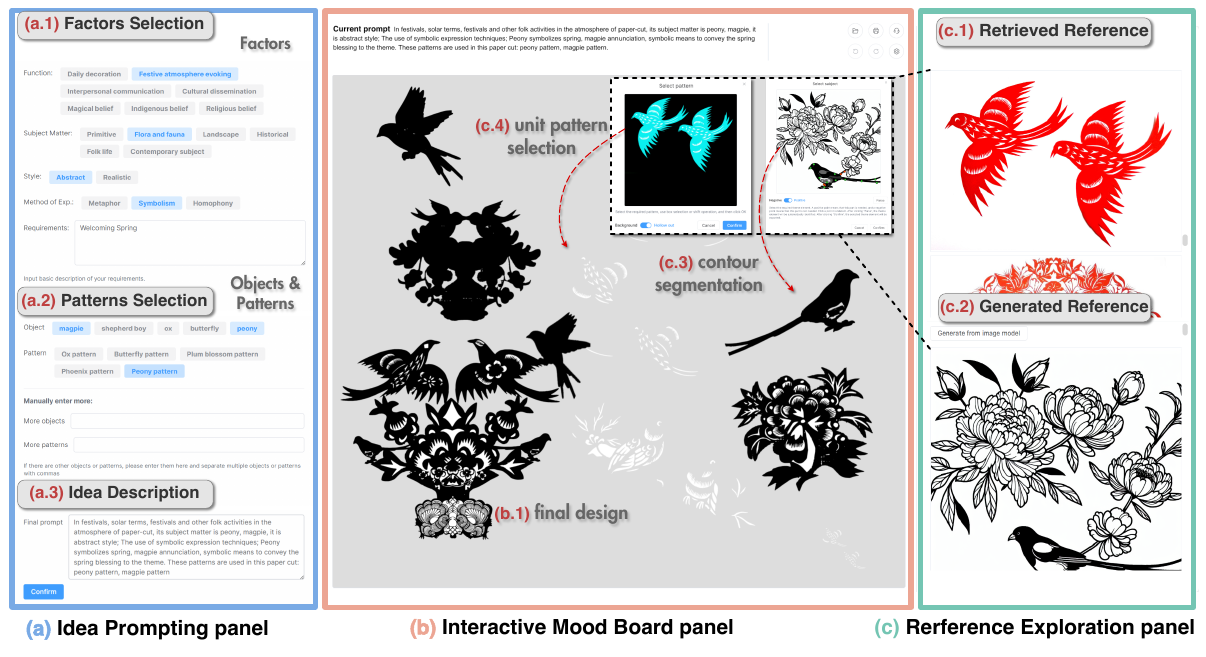}
\caption{\label{figure3}
The interface of HarmonyCut supports user creative paper-cutting design through several panels with guidance and exploration. (a) In the Idea Prompting panel, the system guides users through idea navigation using four factors: function, subject matter, style, and method of expression. It also provides related patterns with interpretations. Users can select suggested content that aligns with their intent or manually edit ideas. (b) The Interactive Mood Board panel allows users to arrange all references, including contours and patterns, for composition. (c) In the Reference Exploration panel, users explore system-suggested content, both retrieved and generated, to select contours and plan pattern layouts, thereby gaining ideas for their paper-cutting designs.}
\Description{This figure shows an interface of HarmonyCut that supports user creative paper-cutting design through several panels through guidance and exploration. (a) In the Idea Prompting panel, the system guides users through idea navigation using four factors: function, subject matter, style, and method of expression. It also provides related patterns with interpretations. Users can select suggested content that aligns with their intent or manually edit ideas. (b) The Interactive Mood Board panel allows users to arrange all references, including contours and patterns, for composition. (c) In the Reference Exploration panel, users explore system-suggested content, both retrieved and generated, to select contours and plan pattern layouts, thereby gaining ideas for their paper-cutting designs.}
\end{figure*}

Drawing from the design space and the derived design goals, we introduced a two-component pipeline (\autoref{figure2}) to guide the development of HarmonyCut (\autoref{figure3}), a GenAI-aided design prototype system. This system can assist users in designing paper-cuttings by allowing them to explore and edit related references, such as knowledge, existing paper-cutting works, and patterns, ensuring alignment with both visual form and cultural meaning. HarmonyCut consists of two main components in the pipeline, corresponding to stages in the GenAI-aided paper-cutting design workflow: ideation (DG1) using LLMs and composition (DG2) employing Text-to-Image models~\autoref{figure2}. Each component is connected to the domain knowledge base we have summarized and supports editing and iteration, thereby enhancing the controllability of the design process (DG3).



\subsection{Ideation Component}
Based on the paper-cutting design space we summarized, which includes taxonomies of ideation factors and the pattern, along with annotation information from 140 works grounded in this design space, we leverage the robust language understanding and reasoning capabilities of GLM-4~\cite{Teamglm:2024:chatglm}, an LLM that supports the construction of specific knowledge bases. By integrating these annotations, 140 paired question-answering prompt templates derived from these annotations, and the design space as domain knowledge for the model's few-shot learning, users can input their preliminary design intent in text and specify the design factor types that meet their needs (\autoref{figure2}). This enables the system to initially provide content recommendations, knowledge, and interpretations derived from both the model and annotations, based on the user's input within the context of our design space. For example, if a user selects the expression method ``\textit{symbolism,}'' HarmonyCut not only suggests relevant objects and patterns but also provides their symbolic meanings, allowing users to verify alignment with their intent. This approach ensures coherence between form and connotation. It not only facilitates the development of users' design ideas but also deepens their understanding of valuable connotative insights. Through a three-step exploration process including factor selection, suggested content, and knowledge exploration, users ultimately form a textual description of ideas.

\subsection{Composition Component}
The composition component aims to guide the user to explore the reference and translate the user's conceptual idea into visual form with a mood board. It encompasses three fundamental stages: related content selection, interested and inspired content arrangement, and cut-out area layout.  All elements in the mood board are presented in SVG format, enabling flexible configuration and editing, including operations such as grouping, flipping, and copy-pasting, which accommodate user preferences and design needs.
\subsubsection{Selection} 
To ensure the quality of paper-cutting references, we employed advanced multi-modal models for retrieval and generating, enabling the system to maintain and reflect the intricate style and thematic relevance of Chinese paper-cutting.
To retrieve related paper-cuttings as references, we used the Chinese version of CLIP~\cite{Yang:2023:chineseclip, Radford:2021:clip}, initially pre-trained on approximately two million text-image pairs. We fine-tuned this model using our annotated dataset of 4,031 text-image pairs, derived from 140 annotated paper-cuttings as detailed in \autoref{sec:content}. Following 30 epochs of training with 80\% of the dataset and validation on the remaining 20\%, the fine-tuned Chinese CLIP achieved recall@1 of 63.46\%, recall@5 of 87.24\%, and recall@10 of 94.93\% on the validation set. These metrics demonstrate that the model effectively maintains the relevance between idea descriptions and paper-cutting references. \revisedtext{To prevent cognitive overload and enhance user navigation within the exploration process, HarmonyCut is designed to retrieve the top 20 most relevant paper-cuttings as references.}
To generate images that are more rational and reflective of the Chinese paper-cutting style using DALL-E-3, prompts can be automatically generated from a carefully crafted template based on object requirements or be edited by the user to include more specific requirements. All paper-cutting works, including both retrieved and generated content, are displayed within the same panel as shown in \autoref{figure3} (c) to facilitate user exploration.

\subsubsection{Arrangement} 
Upon exploring and selecting idea-related references, the composition stage necessitates the precise spatial arrangement of paper-cutting contours. Segment Anything Model (SAM)~\cite{Kirillov:2023:sam}, with its visual prompt points, enables the accurate segmentation of these contours, allowing users to extract content effortlessly. HarmonyCut provides the functionality for users to click on objects and backgrounds, using two labels to prompt segmentation according to their preferences, which is illustrated in the \textit{``contour segmentation''} of \autoref{figure3}. Then, all extracted contours, whether used as design material for combination or as design reference to inspire ideas, are converted into SVG format to facilitate subsequent interactions, and they are compiled within HarmonyCut's central mood board, enabling users to explore various potential combinations of the selected content.

\begin{figure*}[!htbp]
\centering
\includegraphics[width=\textwidth]{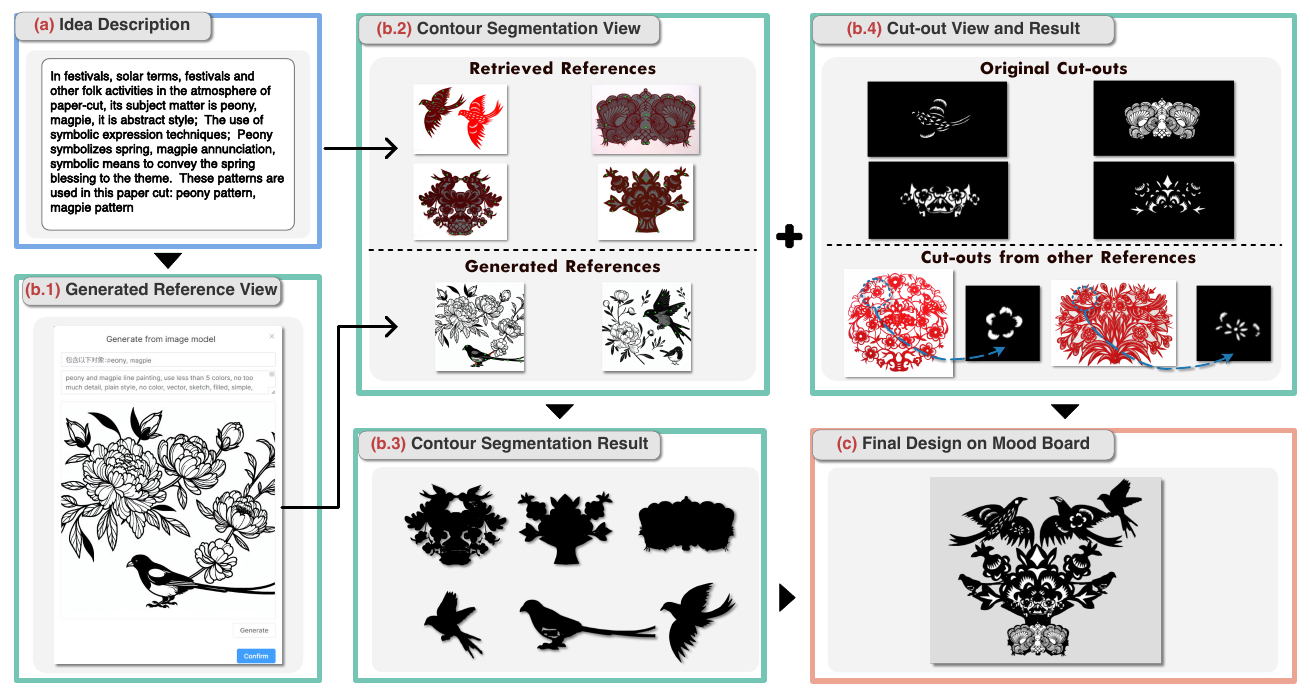}
\caption{\label{figure:detail system}
The detailed process of design with each view and result in HarmonyCut. (a) The idea description from the former ideation. (b.1) The reference generation view based on the prompt; (b.2 and b.3) the contour segmentation view for the selected references and their contours; (b.4) cut-outs from original or other paper cuttings used in the design. (c) The final design is displayed on the mood board.}
\Description{This Figure shows the detailed process of design with each view and result in HarmonyCut. (a) The idea description from the former ideation. (b.1) The reference generation view based on the prompt; (b.2 and b.3) the contour segmentation view for the selected references and their contours; (b.4) cut-outs from original or other paper cuttings used in the design. (c) The final design is displayed on the mood board.}
\end{figure*}

\subsubsection{Cut-out Area} 
As key elements of paper-cutting, these patterns contribute to the unique symbols and styles characteristic of this hollowed-out art form. Each individual varies in the types of patterns used, as well as the density and intricacy in depicting textures, light, and lines. Therefore, the design of pattern layouts is particularly important, as it represents the designer's own unique style.
To effectively plan cut-out areas, it is crucial to extract patterns from paper-cuttings and provide users with insights akin to those offered in the ideation component, where an LLM interprets suggested composite patterns. This requires two primary preprocessing tasks: unit-pattern extraction and classification, which is shown in \autoref{figure2}.
In the preparation of unit patterns for paper-cutting design, we begin by extracting these patterns as masks using OpenCV\footnote{\url{https://github.com/opencv/opencv}}. These masks are then converted into SVG format, which facilitates the subsequent integration of contours and unit patterns. This process is vital for recognizing and employing the basic elements that contribute to composite patterns. Unlike standalone designs, these elements emphasize the content and texture of paper-cutting works. By utilizing SVG-formatted patterns, the system ensures the preservation of the unique symbols and styles inherent to this hollowed-out art form.
To improve pattern recognition and utilization, we fine-tuned the Vision Transformer (ViT)~\cite{Dosovitskiy:2021:vit} with 1,279 annotated unit patterns and labeled all 63,452 unit patterns from 140 paper-cuttings. This resulted in achieving 71\% precision, 60.43\% recall@1, and an F1 score of 63.20\% on the validation dataset, thereby enhancing the model's ability to effectively identify patterns and support the overall design process.


\subsection{User Scenario}
In HarmonyCut, the components aligned with the design workflow and the goals are arranged within a three-panel interface, with the primary mood board centrally positioned, as illustrated in~\autoref{figure3}. \revisedtext{To demonstrate our system, we present a scenario~\cite{Fulton:2000:usagescenario, Kraft:2012:usagescenario} involving a persona, Tom, a paper-cutting enthusiast who is skilled in cutting but less experienced in composition.} Tom uses HarmonyCut to generate ideas and translate them into visual representations for his design task. With the ``Spring Art Exhibition'' approaching at his school, he hopes to leverage HarmonyCut to design a paper-cutting-themed piece called ``Welcoming Spring'' and improve his composition skills.

\subsubsection{Guide User Ideation with Suggestion}
To get some ideas from our system, Tom first input his rough intent of design as ``\textit{paper-cutting with meaning of welcoming spring}'', function as \textit{evoking a festive atmosphere}, subject matter as \textit{flora and fauna}, style as \textit{abstract}, and expression method as \textit{symbolism} to system in~\autoref{figure3} (a.1). HarmonyCut displays objects and patterns related to his intent, including peony, magpie, butterfly, Phoenix, etc. (\autoref{figure3}~(a.2)), as well as the background, structure, common combinations with other patterns, and their corresponding meanings for each recommended pattern. Among the recommended options, Tom still believed that the magpie and peony, two of the most commonly used elements, best suit the design theme. He realized that with this subject, creativity will largely depend on the composition, which is his weak point. Therefore, he prefers to explore others' compositions to guide his design. He then finalized the editable idea description (\autoref{figure3}~(a.3)) and clicks the ``Confirm'' button.

\subsubsection{Exploration on Related Design Reference}
As Tom confirmed his final idea description, the Reference Exploration panel (\autoref{figure3}~(c)) provides two parts of related paper-cutting suggestions, including retrieved and generated paper-cuttings from models.
In \autoref{figure3}~(c.1) view, Tom first found a reference that inspired him on how to structure his composition (two symmetrical birds and a potted flower, as shown in~\autoref{figure2}~(B)). Thus, he explores other magpies and potted flowers as the composition elements. 
Meanwhile, Tom also came across a symmetrical baby chicken paper-cutting \autoref{figure2}~(B) among the suggestions, which inspired him to combine the baby chicken design with the potted flower and magpie paper-cutting. He envisioned using the fresh, lively image of the baby chicken to symbolize the arrival of spring. As a result, this baby chicken paper-cutting was also selected.
Then, Tom browsed paper-cutting images generated by DALL-E-3~(\autoref{figure3}~(c.2) and \autoref{figure:detail system}~(b.1)). Although these images had a strong traditional style and sense of design, Tom had selected an \textit{abstract} style and \textit{paper-cutting image} in the ideation stage, the model kept generating images in traditional Chinese painting or realistic styles, even after he tried adjusting the input prompts. Nevertheless, he found the compositions of these images quite interesting, particularly the idea that “\textit{peonies are blossoming on the back of the magpie.}”
Additionally, Tom felt that the patterns used in the magpie paper-cutting (\autoref{figure3}~(c.4)) he selected—such as the contours, feathers, and eyes—were too simplistic. As a result, he extracted various sawtooth patterns from other retrieved works (\autoref{figure:detail system}~(b.4)).

\subsubsection{Mood Board with Contour and Cut-out}


Based on the references Tom explored in the previous step, he first used SAM by adding visual prompts to segment both the paper-cutting that served as his source of inspiration and the paper-cutting contours he wanted to use as compositional elements~(\autoref{figure:detail system}~(b.2-3)). He placed these onto the mood board canvas. Following the composition of the reference artwork that inspired him, he duplicated and flipped the magpie he selected, symmetrically positioning the two magpies over another potted flower paper-cutting. He then placed the contour of the symmetrical baby chicken paper-cutting, previously selected, at the bottom of the entire design.
For the patterns in the paper-cutting design, Tom drew inspiration from the generated images where flowers blossom on birds. He extracted the sawtooth patterns from other paper-cutting works in~\autoref{figure:detail system}~(b.4), selecting four sawtooth patterns to form a plum blossom pattern, which he placed on the right magpie~(\autoref{figure3}~(c.1)).
Finally, after arranging all the patterns he wanted to add, he completed the final design, shown in~\autoref{figure3}~(b.1).

\subsection{Implementation Details}
Our system utilizes a Flask\footnote{\url{https://flask.palletsprojects.com}}-based back-end, integrated with the GLM-4 APIs for text processing and DALL-E-3 for paper-cutting image generation. The front-end is built using Vite + Vue, with additional support from Element Plus for UI components, Pinia for state management, and Tailwind CSS for styling.
A fine-tuned CLIP model for multi-modal retrieval, ViT-base-16 and RoBERTa-wwm-base are used for visual and text encoding.
For image segmentation, we extended the segment-anything-web UI\footnote{\url{https://github.com/Kingfish404/segment-anything-webui}} framework, which enables extraction and manipulation of image components. The backend supports the use of SAM-ViT-base\footnote{\url{https://github.com/facebookresearch/segment-anything}}~\cite{Kirillov:2023:sam} by default, with flexible configurations for alternative models.
Additionally, a fine-tuned pre-trained ViT (ViT-base-16, pre-trained on ImageNet-21k) is employed for unit-pattern classification.
The frontend supports dynamic SVG manipulation and bitmap-to-SVG conversion using Fabric\footnote{\url{https://fabricjs.com/}} and Potrace\footnote{\url{https://potrace.sourceforge.net/}} libraries. We also implemented custom features such as object dragger and pattern editor, integrated with the overall canvas design interface. \rrtext{The related datasets, comprising 140 paper-cutting images, 4,031 text-image pairs for fine-tuning the multi-modal retrieval model and 1,279 annotated images for fine-tuning the unit pattern recognition model, are detailed in the supplementary material.}

\section{Evaluation}\label{evaluation}
We conducted a within-subjects user study with sixteen participants and an expert evaluation involving three Chinese paper-cutting experts. The objective was to assess HarmonyCut's usability in paper-cutting design and to validate the proposed design workflow. Then, we interviewed participants to evaluate whether HarmonyCut enhances creativity and facilitates the design process. Insights from the expert interviews highlighted how the system addresses challenges in paper-cutting design and identified areas for improvement.
\revisedtext{In contrast, the baseline tool utilized the same GenAI models as HarmonyCut along with their respective official web interfaces of GLM-4\footnote{\url{https://chatglm.cn/}} and DALL-E-3\footnote{\url{https://chatgpt.com/}}. However, It lacked the domain-specific knowledge (i.e., annotated paper-cuttings and fine-tuned models) and the direct editing capabilities incorporated into HarmonyCut. Consequently, users of the baseline tool had to manually input text for design ideation with GLM-4 and generate final designs by crafting their own prompts for DALL-E-3.}

\subsection{Participants}
\revisedtext{We recruited 16 participants (6 females and 10 males; age M=23.94, SD=7.08) through online postings on various social media platforms.} To ensure HarmonyCut could meet design goals from the formative study, participants were selected based on similar criteria for expertise in paper-cutting and GenAI. \revisedtext{The group included 1 paper-cutting master (U7), 2 practitioners (U2, U8), 3 amateurs (U5, U14-U15), and 10 novices (U1, U3-U4, U6, U9-U13, U16). Regarding GenAI expertise, there were 12 knowledgeable users (U1-U6, U9-U10, U13-U16) and 4 novices (U7-U8, U11-U12).} The three Chinese paper-cutting experts (E1-E3) were recruited from social video platforms, each having more than 20, 40, and 20 years of experience in designing, creating, and teaching paper-cutting. They are also recognized as ICH inheritors. \rrtext{The detailed information of the participants is shown in~\autoref{table:evaluation participants} and each participant received 100 CNY (approximately 14 USD).}

\begin{figure*}[!htbp]
\centering
\includegraphics[width=\textwidth]{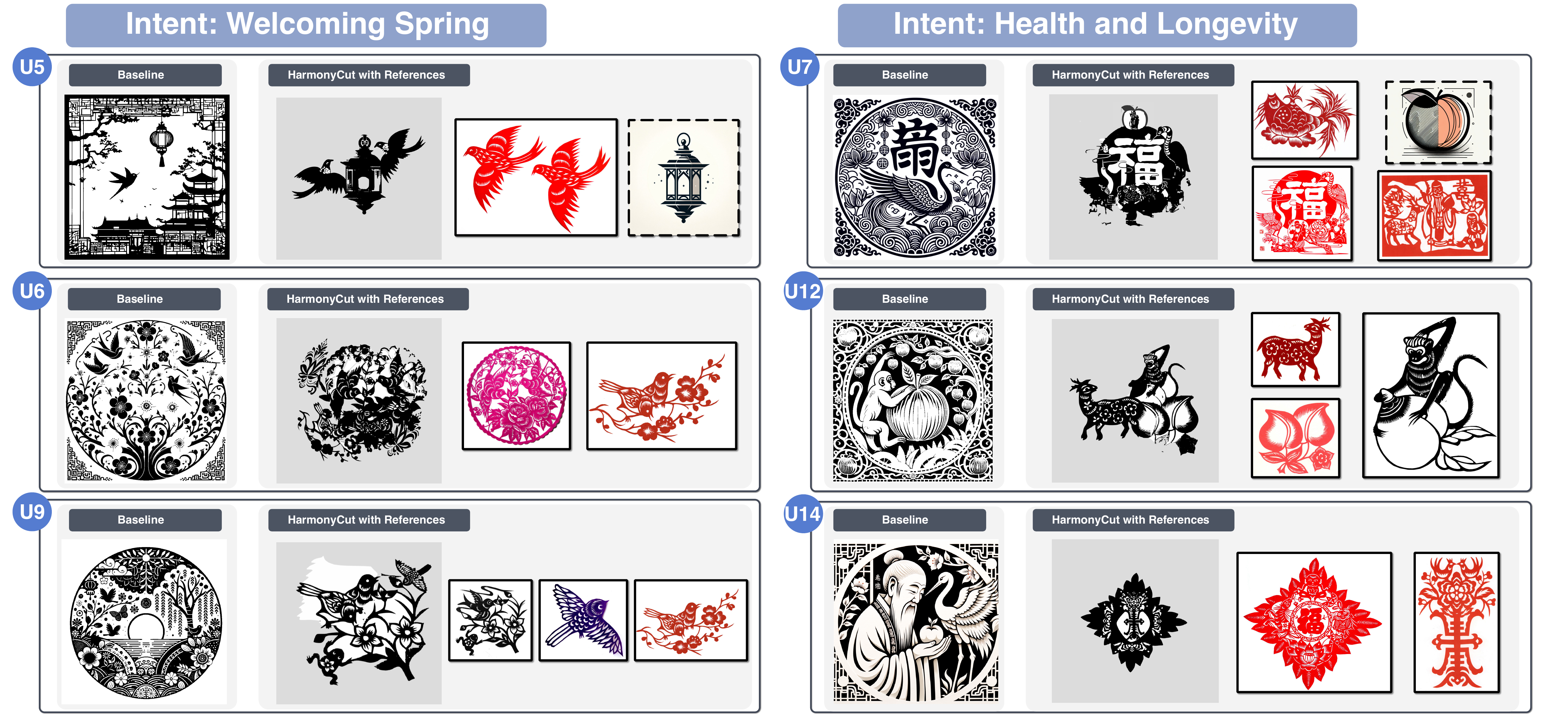}
\caption{\label{figure: user cases}
The 6 paper-cutting design examples were created by the 6 participants in the user study. All solid references are retrieved, while dashed references are generated.}
\Description{This figure shows the 6 paper-cutting design examples created by the participants in the user study. All solid references are retrieved, while ashed references are generated.}
\end{figure*}
\subsection{Procedure and Measures}
In the user study procedure, participants were tasked with designing a paper-cutting that aligned with the provided design themes, such as ``Welcoming Spring'' or ``Health and Longevity.'' Initially, participants are given a 5-minute introduction to the study and its background. For each tool (HarmonyCut or the Baseline), participants underwent three stages: a 15-minute session for tool instruction and task description, a 30-minute period to complete the design task using the respective tool, a 5-minute session to complete a questionnaire, and a 10-minute period for free exploration with the tool. A 5-minute break was scheduled between sessions for each tool. After both experiments, a 15-minute semi-structured interview was conducted. The order of tool usage and tasks was counterbalanced to negate sequence effects on the results. The example outputs generated by participants with varying levels of expertise are presented in~\autoref{figure: user cases}.

To validate the design workflow integrating paper-cutting knowledge with HarmonyCut, participants completed a questionnaire after each session, resulting in two submissions per participant. The questionnaire included three 6-point Likert scale questions focused on design goals: (1) ideation guidance, asking if the tool helps generate ideas; (2) \revisedtext{exploration with reference, asking if the tool supports broad exploration for design, complementary to the Exploration measure in CSI;} and (3) editing flexibility, questioning if the tool allows sufficient editing flexibility. Additionally, the questionnaire incorporated the CSI~\cite{Cherry:2014:csi} (0-10 scale) and NASA-TLX~\cite{Hart:2006:nasa} (0-20 scale) to evaluate usability regarding creative support and perceived workload. 

Prior to the expert interview, each expert was asked to prepare a design intent. During the interview, experts received a 15-minute introduction and system walk-through, including a specific design task. After learning how to use HarmonyCut for the paper-cutting design task, experts were given 30 minutes to create a paper-cutting based on their prepared design intent. After the design session, an interview was conducted on three topics adapted from Xu et al.\cite{Xu:2023:magicalbrush}~(\autoref{table3}), each with two questions to gather feedback.
\begin{table}[!htbp]
\caption{Questions during expert interviews from 3 topics.}
\Description{This table presents the six questions used in the expert interviews, categorized into three topics: culture, creativity, and limitations.}
\label{table3}
\resizebox{0.49\textwidth}{!}{
\renewcommand\arraystretch{1.5}
\begin{tabular}{c|c}
\hline
Topic                        & Question                                                             \\ \hline
\multirow{2}{*}{Culture}     & Does HarmonyCut's suggested content align with its cultural meaning? \\
                             & Can HarmonyCut help the public better understand paper-cutting knowledge? \\ \hline
\multirow{2}{*}{Creativity}  & Does HarmonyCut's guidance limit or enhance your creativity?         \\
                             & What aspects of HarmonyCut inspire new creative ideas?               \\ \hline
\multirow{2}{*}{Improvement} & What are your thoughts on human-AI collaborate design systems?              \\
                             & Where can HarmonyCut be improved?                                    \\ \hline
\end{tabular}
}
\end{table}

\subsection{Results}
Based on the qualitative and quantitative data collected from both studies, we found that HarmonyCut effectively facilitates the generation of creative ideas through guided support. It aided participants in exploring valuable related references and allowed them to incorporate their own creative ideas into the paper-cutting design process. 

\begin{figure*}[!htbp]
\centering
\includegraphics[width=\textwidth]{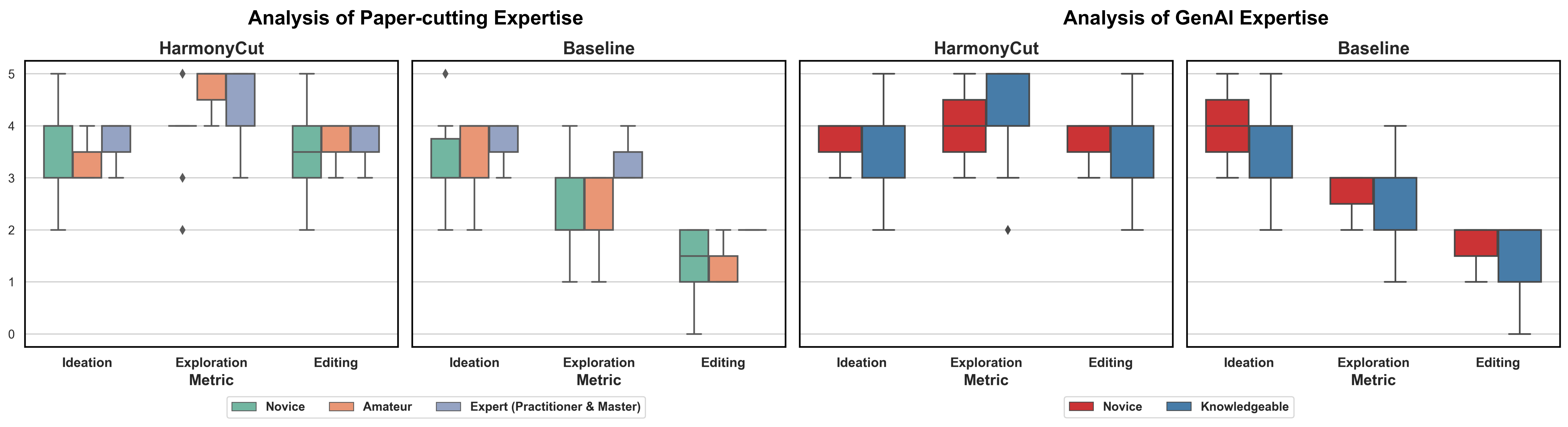}
\caption{\label{figure:DG}
Sixteen participants ratings the design goals questionnaire across different expertise levels on paper-cutting and GenAI.}
\Description{This figure shows the sixteen participants' ratings to the design goals questionnaire across different expertise levels on paper-cutting and GenAI.}
\end{figure*}

\begin{figure*}[!htbp]
\centering
\includegraphics[width=\textwidth]{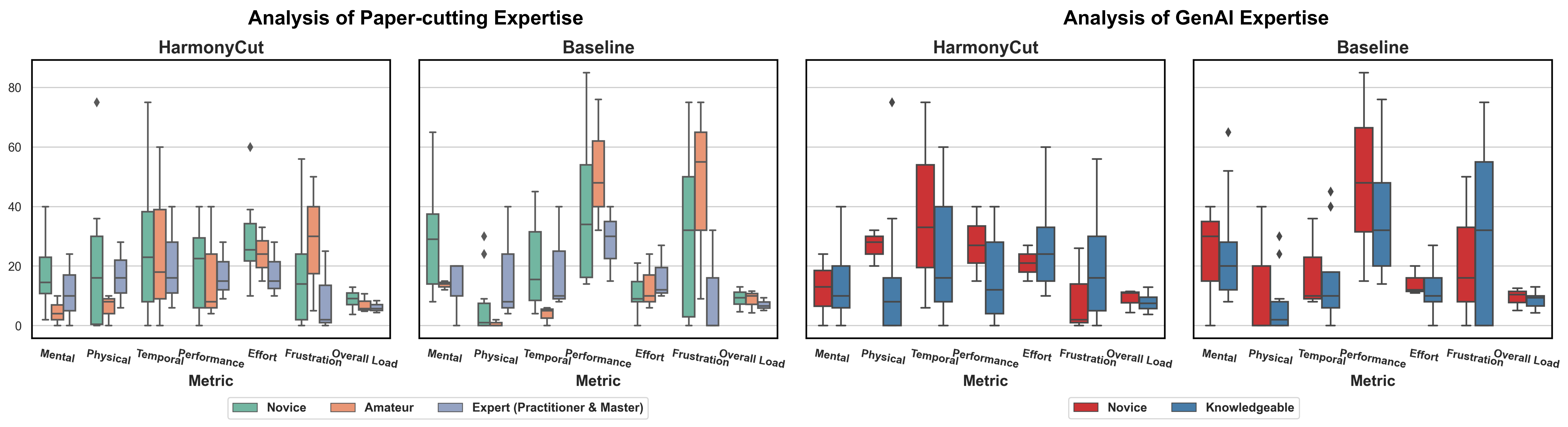}
\caption{\label{figure:NASA}
Sixteen participants ratings to the NASA-TLX perceived load questionnaire across different expertise levels on paper-cutting and GenAI.}
\Description{This figure shows the sixteen participants' ratings to the NASA-TLX perceived load questionnaire across different expertise levels on paper-cutting and GenAI.}
\end{figure*}

\begin{figure*}[!htbp]
\centering
\includegraphics[width=\textwidth]{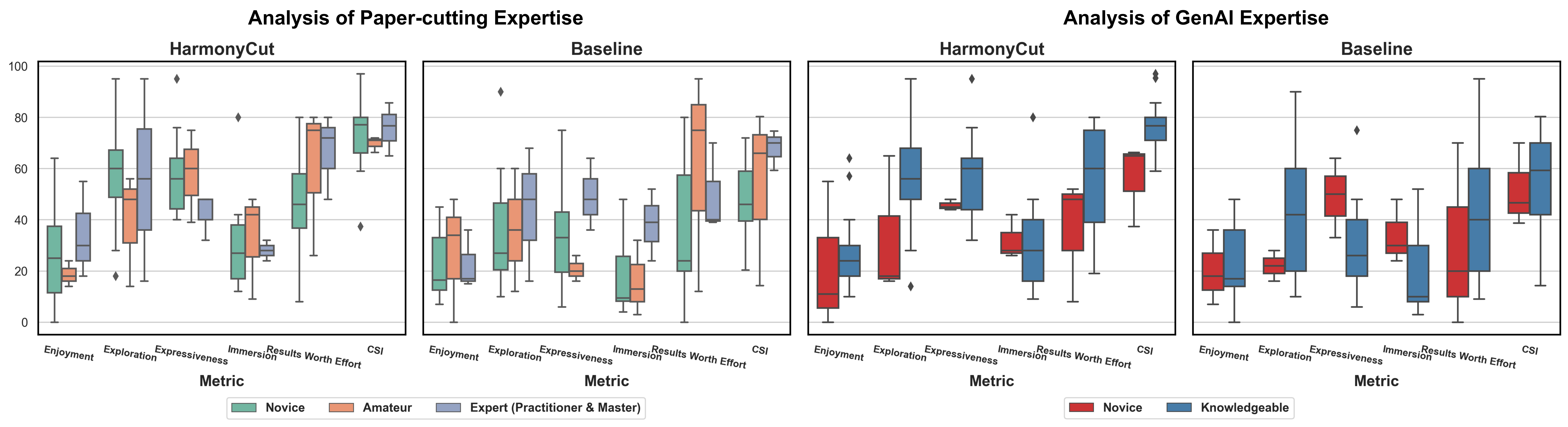}
\caption{\label{figure:CSI}
Sixteen participants ratings on the Creative Support Index questionnaire across expertise levels in paper-cutting and GenAI.}
\Description{This figure shows the sixteen participants' ratings to the Creative Support Index questionnaire across different expertise levels on paper-cutting and GenAI.}
\end{figure*}

To assess whether each design goal was achieved within HarmonyCut and whether the system effectively addressed the challenges associated with these goals in the design process, we first analyzed the questionnaire data in the user study. Considering the small sample size (N=16) and the ordinal nature of the data, we employed a Wilcoxon signed-rank test to compare the differences between HarmonyCut and the baseline tool. \revisedtext{Although~\autoref{table4} showed that the average performance across most metrics is better than the Baseline, participants had varying levels of expertise. Therefore, we also analyzed their opinions of the two tools based on their paper-cutting and GenAI expertise levels~(in \autoref{figure:DG},\ref{figure:NASA},\ref{figure:CSI}).} Subsequently, we gathered feedback from the user study and the expert interview.

\begin{table*}[!htbp]
\caption{Survey results of participant opinion about design goals, NASA-TLX questionnaire, and Creativity Support Index.}
\Description{This table illustrates the results from the survey on participants' opinions about the design goals, the NASA-TLX questionnaire, and the Creativity Support Index.}
\label{table4}
\renewcommand\arraystretch{1.2}
\resizebox{0.95\textwidth}{!}{
\begin{tabular}{ccccccc}
\hline
\multicolumn{2}{c|}{\multirow{2}{*}{Indicator}}                                                                  & \multicolumn{2}{c|}{\textbf{HarmonyCut}}                  & \multicolumn{2}{c|}{\textbf{Baseline}}                    & \multirow{2}{*}{P} \\ \cline{3-6}
\multicolumn{2}{c|}{}                                                                                            & \multicolumn{1}{c|}{Mean}   & \multicolumn{1}{c|}{SD}     & \multicolumn{1}{c|}{Mean}   & \multicolumn{1}{c|}{SD}     &                    \\ \hline
\multicolumn{1}{c|}{\multirow{3}{*}{Survey related to design goals}} & \multicolumn{1}{c|}{Ideation}             & \multicolumn{1}{c|}{3.563}  & \multicolumn{1}{c|}{0.727}  & \multicolumn{1}{c|}{3.375}  & \multicolumn{1}{c|}{0.957}  & 0.472              \\ \cline{2-7} 
\multicolumn{1}{c|}{}                                                & \multicolumn{1}{c|}{Exploration}          & \multicolumn{1}{c|}{4.125}  & \multicolumn{1}{c|}{0.885}  & \multicolumn{1}{c|}{2.563}  & \multicolumn{1}{c|}{0.892}  & 0.0029**           \\ \cline{2-7} 
\multicolumn{1}{c|}{}                                                & \multicolumn{1}{c|}{Editing}              & \multicolumn{1}{c|}{3.625}  & \multicolumn{1}{c|}{0.806}  & \multicolumn{1}{c|}{1.438}  & \multicolumn{1}{c|}{0.727}  & 0.00003***         \\ \hline
\multicolumn{1}{c|}{\multirow{7}{*}{NASA-TLX}}                       & \multicolumn{1}{c|}{Mental}               & \multicolumn{1}{c|}{14.250} & \multicolumn{1}{c|}{11.958} & \multicolumn{1}{c|}{23.625} & \multicolumn{1}{c|}{16.931} & 0.026*             \\ \cline{2-7} 
\multicolumn{1}{c|}{}                                                & \multicolumn{1}{c|}{Physical}             & \multicolumn{1}{c|}{16.813} & \multicolumn{1}{c|}{19.641} & \multicolumn{1}{c|}{7.625}  & \multicolumn{1}{c|}{12.447} & 0.046*             \\ \cline{2-7} 
\multicolumn{1}{c|}{}                                                & \multicolumn{1}{c|}{Temporal}             & \multicolumn{1}{c|}{24.625} & \multicolumn{1}{c|}{22.102} & \multicolumn{1}{c|}{16.688} & \multicolumn{1}{c|}{14.858} & 0.460              \\ \cline{2-7} 
\multicolumn{1}{c|}{}                                                & \multicolumn{1}{c|}{Performance}          & \multicolumn{1}{c|}{18.875} & \multicolumn{1}{c|}{13.837} & \multicolumn{1}{c|}{39.750} & \multicolumn{1}{c|}{23.345} & 0.0041**           \\ \cline{2-7} 
\multicolumn{1}{c|}{}                                                & \multicolumn{1}{c|}{Effort}               & \multicolumn{1}{c|}{25.438} & \multicolumn{1}{c|}{13.008} & \multicolumn{1}{c|}{12.188} & \multicolumn{1}{c|}{7.458}  & 0.0045**           \\ \cline{2-7} 
\multicolumn{1}{c|}{}                                                & \multicolumn{1}{c|}{Frustration}          & \multicolumn{1}{c|}{18.625} & \multicolumn{1}{c|}{19.328} & \multicolumn{1}{c|}{30.125} & \multicolumn{1}{c|}{28.477} & 0.157              \\ \cline{2-7} 
\multicolumn{1}{c|}{}                                                & \multicolumn{1}{c|}{Overall Load}         & \multicolumn{1}{c|}{7.908}  & \multicolumn{1}{c|}{2.980}  & \multicolumn{1}{c|}{8.667}  & \multicolumn{1}{c|}{2.805}  & 0.562              \\ \hline
\multicolumn{1}{c|}{\multirow{6}{*}{Creativity Support Index}}       & \multicolumn{1}{c|}{Enjoyment}            & \multicolumn{1}{c|}{27.125} & \multicolumn{1}{c|}{18.395} & \multicolumn{1}{c|}{23.438} & \multicolumn{1}{c|}{14.660} & 0.562              \\ \cline{2-7} 
\multicolumn{1}{c|}{}                                                & \multicolumn{1}{c|}{Exploration}          & \multicolumn{1}{c|}{54.250} & \multicolumn{1}{c|}{26.055} & \multicolumn{1}{c|}{37.563} & \multicolumn{1}{c|}{23.218} & 0.073              \\ \cline{2-7} 
\multicolumn{1}{c|}{}                                                & \multicolumn{1}{c|}{Expressiveness}       & \multicolumn{1}{c|}{55.375} & \multicolumn{1}{c|}{16.552} & \multicolumn{1}{c|}{34.250} & \multicolumn{1}{c|}{18.724} & 0.0076**            \\ \cline{2-7} 
\multicolumn{1}{c|}{}                                                & \multicolumn{1}{c|}{Immersion}            & \multicolumn{1}{c|}{30.688} & \multicolumn{1}{c|}{17.621} & \multicolumn{1}{c|}{20.750} & \multicolumn{1}{c|}{15.902} & 0.074              \\ \cline{2-7} 
\multicolumn{1}{c|}{}                                                & \multicolumn{1}{c|}{Results Worth Effort} & \multicolumn{1}{c|}{52.313} & \multicolumn{1}{c|}{22.934} & \multicolumn{1}{c|}{42.375} & \multicolumn{1}{c|}{28.577} & 0.222              \\ \cline{2-7} 
\multicolumn{1}{c|}{}                                                & \multicolumn{1}{c|}{CSI}                  & \multicolumn{1}{c|}{73.250}  & \multicolumn{1}{c|}{14.283} & \multicolumn{1}{c|}{52.792} & \multicolumn{1}{c|}{19.734} & 0.0013**           \\ \hline
\multicolumn{7}{l}{*p<0.05; **p<0.01; ***p<0.001}                                   
\end{tabular}
}
\end{table*}

\subsubsection{Ideation with Guidance}
Regarding the first design goal, factor-oriented guidance for ideation, as shown in~\autoref{table4}, while the results for ideation were not statistically significant, the average scores indicated that participants perceived HarmonyCut as providing better support for their \textbf{Ideation} process (HarmonyCut: M=3.563, SD=0.727 / Baseline: M=3.375, SD=0.957). Additionally, the NASA-TLX results revealed that participants experienced a significantly lower \textbf{Mental} workload when using HarmonyCut (M=14.250, SD=11.958 / Baseline: M=23.625, SD=16.931 / P=0.026*, W=21.0) compared to the Baseline. This suggests that HarmonyCut's guided ideation, facilitated by factor-based options, effectively supported users during the ideation stage, especially for tasks requiring mental effort.

Furthermore, participants noted that the baseline tool tended to recommend overly broad information or required iteration, making it challenging to decide on an idea. As U3 mentioned, ``\textit{I gave a vague description and had to spend time filtering suggestions.}'' In contrast, HarmonyCut's selection of factors and suggested content helped users understand paper-cutting knowledge through introductions and interpretations. U5 highlighted this advantage, stating, ``\textit{HarmonyCut provided relevant patterns, each with explanations of their meaning and how they fit the design purpose.}''  Meanwhile, in the expert interview, E1’s responses to questions 1 and 2 on the cultural topic, given E1's many years of experience teaching paper-cutting, align with U5’s points: ``\textit{Aligning suggested form and meaning is not enough for understanding. It is through interpretation that users can really grasp the content and, in turn, find inspiration during the learning process.}''. This guidance aided users in the ideation process, helping them quickly construct initial ideas and comprehend the rationale behind their choices, as U3 concluded, ``\textit{This helped me quickly construct initial ideas and understand their choices.}''

In addition, to evaluate how DG1 addresses C2 in~\autoref{sec:formative}, we collected expert feedback on creativity and culture during the interviews. All three experts agreed that the factors-oriented guidance did not limit their creativity. They also acknowledged that a system supported by domain knowledge, which enhances understanding through explanations, was highly beneficial for learning about paper-cutting traditions from different regions and cultures, aiding in innovative ideation.

\subsubsection{Reference Exploration}
\revisedtext{The results indicated that HarmonyCut significantly enhances the exploration process, as reflected in the \textbf{Exploration} measure in design goal questionnaire (HarmonyCut: M=4.125, SD=0.885 / Baseline: M=2.563, SD=0.892 / p=0.0029**, W=9.0). Besides, HarmonyCut's \textbf{Exploration} score in the CSI was higher than that of the baseline tool (HarmonyCut: M=54.250, SD=26.055 / Baseline: M=37.563, SD=23.218). These results suggested that HarmonyCut supports a more diverse and creative exploration compared to the Baseline.}

Several participants provided insightful feedback. U4 and U8 both noted that the ``\textit{Baseline system tended to return similar results, regardless of input changes, limiting creative exploration.}'' In contrast, U5 highlighted that ``\textit{HarmonyCut's combination of design retrieval and idea-based generation allowed for more efficient exploration.}'' This approach helped avoid the pitfalls of inconsistent generative results while offering a broader creative scope than relying solely on retrievable data.
To analyze based on participants' varying levels of expertise, we found that U7, a master of paper-cutting~(\autoref{figure:DG},\autoref{figure:CSI}), felt that the exploration feature offered the same capability as the Baseline (\textbf{Exploration} in Design Goal: HarmonyCut=4, Baseline=4; \textbf{Exploration} in CSI: HarmonyCut=16, Baseline=16), stating: ``\textit{I've seen quite a lot of paper-cuttings. The additional retrieval function in HarmonyCut did not significantly enhance the sense of exploration.}'' Interestingly, U7 was recently invited to design a government-themed paper-cutting piece. Seeking inspiration, U7 reviewed realistic paper-cuttings of steel factories retrieved by HarmonyCut in the free exploration session. Using these as references, along with HarmonyCut's generative model to explore common themes related to steel factories, U7 noted that ``\textit{Although the exploration content is limited due to the restricted dataset, reference-based exploration did help me with composition. Also, using the system saved me a lot of time compared to sketching, while the Baseline was almost useless for planning the cut-out layout.}'' (\textbf{Temporal Load}: HarmonyCut=6, Baseline=10). For participants with varying expertise levels in both paper-cutting and GenAI, HarmonyCut was generally perceived to enhance exploration capabilities, as illustrated in~\autoref{figure:DG} and~\autoref{figure:CSI}.

\subsubsection{Editing with Effort}\label{sec:editing}
For content editing, HarmonyCut supports segmentation based on the contours of composite patterns, extraction of unit patterns, and retrieval of related works. In contrast, the baseline tool only allows output adjustments through input modifications. The results indicate that participants generally perceived HarmonyCut as offering more flexible \textbf{Editing} capabilities compared to the baseline, enhancing controllability (HarmonyCut: M=3.625, SD=0.806 / Baseline: M=1.438, SD=0.727 / p=0.00003***, W=0) through user interaction rather than relying solely on model outputs. \revisedtext{Furthermore, although the results for \textbf{Enjoyment} were not statistically significant (HarmonyCut: M=27.125, SD=18.395 / Baseline: M=23.438, SD=14.660), some participants expressed enjoyment in editing their designs using our system. Moreover, despite being knowledgeable about GenAI and using prompt engineering to enhance outputs, some participants argued that the lack of control in the baseline tool hindered design. Furthermore, some participants attributed the improved ability to \textbf{Express} ideas with HarmonyCut (HarmonyCut: M=55.375, SD=16.552; Baseline: M=34.250, SD=18.724; p=0.0076**, W=13.0), compared to the Baseline, to its controllable editing support. For instance, U6 remarked, ``\textit{The inability to directly edit the output and the uncontrollable reasoning process made it difficult for me to enjoy the design experience.}'' U14 also mentioned about ``\textit{Especially for content related to Chinese characters, the Baseline often fails to meet the requirements, generating some meaningless stroke combinations. By comparison, I still prefer a controllable design process to express my idea.}''} To the expert interview, E3 noted that ``\textit{the ability to edit undesirable parts of the generated image is a critical function in HarmonyCut, improved the controllability issue of GenAI.}''

\subsubsection{\revisedtext{Design Process and Design Performance}}\label{sec:process performance}
\revisedtext{Chinese paper-cutting embodies a fundamental connotation: the aspiration for a better life. This connotation is directly expressed through the selection of content and composition, which serve as key considerations in the creative process~\cite{Lin:1974:howtopapercutting, Zhang:2021:papercuttingteaching, Li:1998:monopapercutting}. Experienced paper-cutting educators (E1-E3) also consistently emphasized the importance of training students to choose appropriate content and develop effective compositions as a critical aspect of learning paper-cutting. Consequently, integrating support for these aspects into the design process is essential. 
This was partially reflected in the \textbf{Performance} (HarmonyCut: M=18.875, SD=13.837 / Baseline: M=39.750, SD=23.345 / p=0.0041**, W=9.5) cross users with different levels of expertise. Experts recognized HarmonyCut in~\autoref{figure:NASA} as effective in supporting reference selection and composition. Novices acknowledged the system's \textbf{Performance}~(\autoref{figure:NASA}) and additionally reported a more \textbf{Immersive} experience~(\autoref{figure:CSI}) during the design process. Notably, U3, U9, and U13, as young novices, emphasized that the design process in HarmonyCut not only helped them learn the knowledge of paper-cutting but also sparked a deeper interest in the craft. Such knowledge sharing and interest of the young generation play a role in influencing the preservation and transmission of ICH~\cite{Affleck:2008:newgeneration, Mancacaritadipura:2009:newgeneration}, like paper-cutting. Nevertheless, beyond initial interest, long-term accumulation and profound knowledge are necessary to foster creativity and innovation, ensuring the inheritance and vitality of ICH, as also discussed in~\autoref{limitations}. }

\subsubsection{System Limitations}\label{limitations}
Based on feedback from participants in the user study and expert interviews, we have summarized the current limitations of the system, which are also discussed in~\autoref{lf} with future directions:

\noindent\textbf{Cultural Knowledge Diversity:} \revisedtext{While the current design space is relatively comprehensive, given the diversity of cultural knowledge, especially in profound folk arts like paper-cutting. Thus, there is room for further expansion of labeled data, including the taxonomy and recommended reference (proposed by E1).} Continuously enriching the dataset will ensure the design space remains reasonable and that the paper-cutting knowledge and explanations provided by the model are more accurate. This is an ambitious and ongoing effort. U7 made a similar suggestion, pointing out that the collected and annotated paper-cutting data still needs to be expanded, as the bottleneck appears to lie in the data. Expanding this dataset would improve the usability of HarmonyCut.

\noindent\textbf{Effort-Intensive Design Process:} Although the contour + pattern approach enhances the flexibility of the design process, it requires more effort compared to simply inputting text and receiving generated images (U1-U2, U4, U10-11). Users need to invest time in image segmentation and pattern selection. This trade-off is reflected in the NASA-TLX results, where HarmonyCut shows a lower \textbf{Mental} load but significantly higher \textbf{Physical} load and \textbf{Effort} compared to the baseline~(\autoref{table3}). However, balancing these factors is essential.

\noindent\textbf{\revisedtext{Imitation and Creation in Design:}} \revisedtext{
Although HarmonyCut is considered significantly more expressive in~\autoref{sec:editing}, U7 and U8~(\autoref{figure:CSI}) found its expressiveness limited, attributing this to the restricted reference list and lack of more types of editing tools, which they felt constrained their expertise. Besides, some novice users, on the other hand, overly relied on the recommended references for their designs, leading to imitation issues. From another perspective, E1, E2, and Lin~\cite{Lin:1974:howtopapercutting} emphasized that reference-based imitation is, to some extent, a necessary step in the creative process. In their teaching practices, students were required to engage in extensive imitation to acquire relevant knowledge, develop skills, and draw inspiration~(\autoref{sec:process performance}), which was align with the view of Okada et al.~\cite{Okada:2017:imitation}.
Paper-cutting design should extend beyond imitation, which serves as a foundational step in learning the craft. To foster creativity, it is essential to integrate new objects and meanings into the design process. Achieving this requires time, training, and experience, moving past the mere reuse of references to maintain creativity and drive innovation.} 

\section{Discussion}

We introduce the design space and workflow of paper-cutting and propose a novel GenAI-aided creativity support tool, HarmonyCut, which supports reference-guided exploration from ideation to composition of paper-cutting design. Based on our findings, we suggest some design implications for future creativity support tools.


\subsection{Guidance and Reference in Different Context}\label{discussion:guidance reference}
As reflected in the findings and expert feedback from the evaluation (\autoref{evaluation}), users' preferences and assessments of the system's features were influenced by their varying levels of expertise across different fields. Experts tended to favor more exploration, while novices, due to their lack of domain knowledge, preferred the system to make recommendations for them. Therefore, it is crucial to consider the differing workflows of users with varying expertise to support design. This ensures that users in different contexts can effectively use the system to support their design tasks. This approach is also supported by findings from previous research~\cite{Xiao:2024:typedance, Zhou:2023:filtererink, Yan:2022:flatmagic}.

In the ideation process, experts typically have a clear design intent and seek comprehensive references to inspire creativity and refine their ideas. On the other hand, novices often struggle to effectively translate tasks into concrete ideas, making them more reliant on the system to guide them through the ideation process.
Additionally, we observed that novices tend to imitate references rather than use them for creative ideation. Therefore, the system needs to provide explanations alongside the guidance to help them understand the underlying knowledge and design with that understanding. 
\revisedtext{For experts, the focus is on providing references that are comprehensive to better support their advanced creative needs and ensure expressiveness. Due to the differing priorities of novice and expert users, experts may perceive GenAI as inadvertently constraining the creative design process, as also discussed in~\autoref{limitations}. This constraint arises from the nature of GenAI-aided guidance and references, which are inherently linked to the dataset~\cite{Cui:2024:chatlaw, Wang:2023:methodsknowledge} and may restrict creative outcomes to the dataset domain. U2 emphasized that paper-cutting, as a highly abstract form of expression, depends on distinctive personal styles to achieve creative results. However, outputs from GenAI with stereotypes may lead to stylistic homogenization. This not only limits the diversity of creative expression but also raises concerns about copyright~\cite{Samuelson:2023:copyright, Bianchi:2023:stereotype, Zhou:2024:biasgenerativeai}.}

Without proper guidance, novices can easily get lost in many suggestions, while experts can draw inspiration from more abstract references. For novices, more specific guidance helps them extract useful suggestions and apply them to their designs.
In addition to the guidance and references provided in the system workflow, there are trade-offs in the controllabliltiy. On one hand, as an end-to-end model, GenAI results cannot be directly edited, requiring users to iteratively adjust their input, which limits controllability. On the other hand, our experiments revealed that while the system significantly improves editability and control, it also increases the physical and cognitive demands on users compared to simple text-based interaction. Therefore, future research should focus on finding a balance in GenAI-aided design and improving interaction methods to reduce these additional burdens while ensuring adequate control remains in the creative process.


\subsection{Integrate the Cultural Knowledge with GenAI Support Creative Design}
GenAIs have demonstrated strong capabilities in both language and visual understanding, allowing users to engage in the creative process through natural language input. However, in tasks involving abstract cultural and artistic design, these models often fail to fully capture user intent, resulting in outputs that may not meet expectations. The findings from this study, alongside insights from previous research~\cite{Messer:2024:cocreating, Garcia:2024:paradox, Chung:2023:artinter}, highlight the limitations of GenAI in comprehending and interpreting cultural elements. To prevent these shortcomings from impeding creative support, it is essential to incorporate cultural knowledge into the models. Although large-scale data augmentation offers one possible solution, it is limited by both the scale and complexity of data collection.
E1 emphasized that the taxonomy used to organize knowledge through design factors effectively captures a significant portion of paper-cutting knowledge. 

However, the vast and intricate nature of cultural knowledge makes it impractical to rely solely on dataset expansion to improve GenAI's ability to support culturally relevant design. A more effective approach may involve structuring cultural content. Previous work in the field of painting~\cite{Chung:2023:promptpaint} has attempted to extract vague concepts and process them individually, but these models often lack the abstract knowledge necessary, resulting in a disconnect between abstract ideas and concrete visual outputs. By structuring abstract content into design tasks and design spaces, models can generate more appropriate outputs for cultural art-related tasks.

In this study, HarmonyCut organizes the paper-cutting design process around factors and patterns, while Magical Brush uses symbols as fundamental elements in Chinese painting~\cite{Xu:2023:magicalbrush}. Similarly, template-based methods have been applied to dynamic shadow puppetry creation~\cite{Yao:2024:shadowmaker}. \revisedtext{These structured approaches provide a foundational framework for complex and profound cultural knowledge, allowing models to better understand with cultural content, such as the visual selection and composition in HarmonyCut, partially bridging gaps in expertise and cultural background. This makes it possible for public users to access relevant knowledge, participate in cultural creative design, and even further develop expertise, as supported by the feedback from E1–E3 agreeing with the cultural aspects in~\autoref{table3}, thereby engaging and supporting cultural communication while sustaining the vitality of cultural content such as ICH.}
\rrtext{While our structured approach, which relies on manual annotation and model fine-tuning, enhances the model's ability to understand cultural knowledge, it remains limited in achieving comprehensive coverage of large-scale knowledge, as further discussed in~\autoref{lf}.}

\subsection{Limitations and Future Work}\label{lf}
Our work has several limitations that future work can address.
First, the system currently does not support collaborative creation, although prior work has demonstrated that sharing mood boards~\cite{Koch:2019:mayai, Koch:2020:semanticcollage} can enhance the design process. Future work could explore GenAI-aided creativity in multi-user collaboration by implementing features that facilitate cooperative design.
\revisedtext{Second, despite efforts to recruit users with varying levels of familiarity with GenAI, the limited sample size in both formative and user studies resulted in insufficient diversity, affecting the generalizability of the design process. To enhance the system's adaptability, we plan to expand the scale of formative studies to validate and refine the design space and broaden user studies to include participants with more diverse backgrounds and expertise levels. This will help investigate how GenAI expertise influences the use of Creative Support Tools in design.
While the current design space is relatively comprehensive according to the evaluation results and feedback, as a folk art with a vast and profound nature, paper-cutting requires further expansion of labeled data to support a more exhaustive reference list. To ensure broader applicability of the design space, we will continue to update and expand the dataset to support more extensive and in-depth research into paper-cutting art.}
Additionally, the prototype system supports only 2D monochrome paper-cutting, which is the most prevalent form of Chinese paper-cutting. However, it does not accommodate multicolored paper-cutting.
Our current system allows users to transition from their initial intent through ideation to finalizing the composition, ultimately resulting in the design of a paper-cutting piece. Creators can use this design directly as a guide to complete the final creation with knives or scissors. This advancement provides an opportunity to further explore the fabrication aspect, focusing on how tools can support the complete paper-cutting creation process, from design to physical realization.

\section{Conclusion}

This work systematically investigates the workflow and design space of paper-cutting with GenAI assistance. The research prototype, HarmonyCut, employs factor-oriented guidance for ideation and reference-based exploration for composition to address the challenges inherent in designing a paper-cutting. Our user study and expert interviews demonstrated that HarmonyCut enhances users' understanding of paper-cutting knowledge and assists in translating intent into ideas, as well as facilitating the rich exploration of references for paper-cutting composition, resulting in greater engagement and performance compared to the baseline. Feedback from participants validated the comprehension of the workflow and design space and highlighted future directions for improvement. We are eager to further extend the concepts within HarmonyCut to support the traditional art design process with GenAI assistance.

\begin{acks}
This research was partially supported by the National Natural Science Foundation of China (No. 62202217), Guangdong Basic and Applied Basic Research Foundation (No. 2023A1515012889), and Guangdong Key Program (No. 2021QN02X794). We thank all of our study participants for their insightful discussions, constructive feedback, and the dedication to the preservation and dissemination of paper-cutting.
We acknowledge the partial use of a large language model (LLM), specifically ChatGPT, to assist in the writing process. The LLM was employed as a tool for polishing the manuscript to enhance the clarity and quality of the text.
\end{acks}

\balance
\bibliographystyle{ACM-Reference-Format}
\bibliography{reference}
\clearpage
\appendix
\section{\revisedtext{Formative Study}}\label{A:formative participants}

\subsection{\rrtext{Expertise Levels of Paper-cutting}}\label{expert clarification}
\rrtext{As mentioned in~\autoref{A:formative participants}, the key criterion for distinguishing experts (i.e., master and practitioner) is whether they have undergone systematic training (specifically, apprenticing under an inheritor) or have transitioned from amateur to experts status through the accumulation of substantial paper-cutting experience and even been finally recognized as an inheritor. In our interviews, P1-P4 highlighted that they developed an interest in paper-cutting as early as 3-4 years old. Their families subsequently petitioned inheritors to accept these children as apprentices, and the acceptance of such requests indicated the masters' recognition of their talent. As a result, they began receiving systematic and professional training by the age of 5-9. This practice of training young children is relatively common in the Chinese paper-cutting inheritance. However, because this tradition relies on early engagement, the decreasing interest among the younger generation has partially contributed to the challenges faced in the inheritance of ICH, including paper-cutting.}

\begin{table}[H]
\caption{\revisedtext{Summary of participants interviewed in formative study.}}
  \Description{This table demonstrates the summary of participants interviewed in the formative study.}
  \label{table:formative participants}
\resizebox{0.49\textwidth}{!}{
\renewcommand\arraystretch{1.4}
\begin{tabular}{ccccccc}
\hline
ID & Sex    & Age & Paper-cutting Expertise & GenAI Expertise     & Location      & Platform \\ \hline
P1 & Male   & 30  & Master (21 years)       & Novice             & Central China & Bilibili \\
P2 & Female & 28  & Practitioner (18 years) & Knowledgeable User             & East China    & Bilibili \\
P3 & Female & 24  & Practitioner (18 years) & Novice             & Southwest     & Bilibili \\
P4 & Male   & 49  & Master (40 years)       & Novice             & Northwest     & Douyin   \\
P5 & Female & 59  & Master (40+ years)      & Novice             & Northeast     & Douyin   \\
P6 & Male   & 25  & Novice                  & Professional       & North China   & WeChat   \\
P7 & Male   & 26  & Amateur (3 years)       & Knowledgeable User & Central China & WeChat   \\ \hline
\end{tabular}
}
\end{table}

\section{Content Analysis}

\subsection{\revisedtext{Information of 140 Sampled Images}}\label{A:sample filter}
\begin{figure}[H]
\centering
\includegraphics[width=0.49\textwidth]{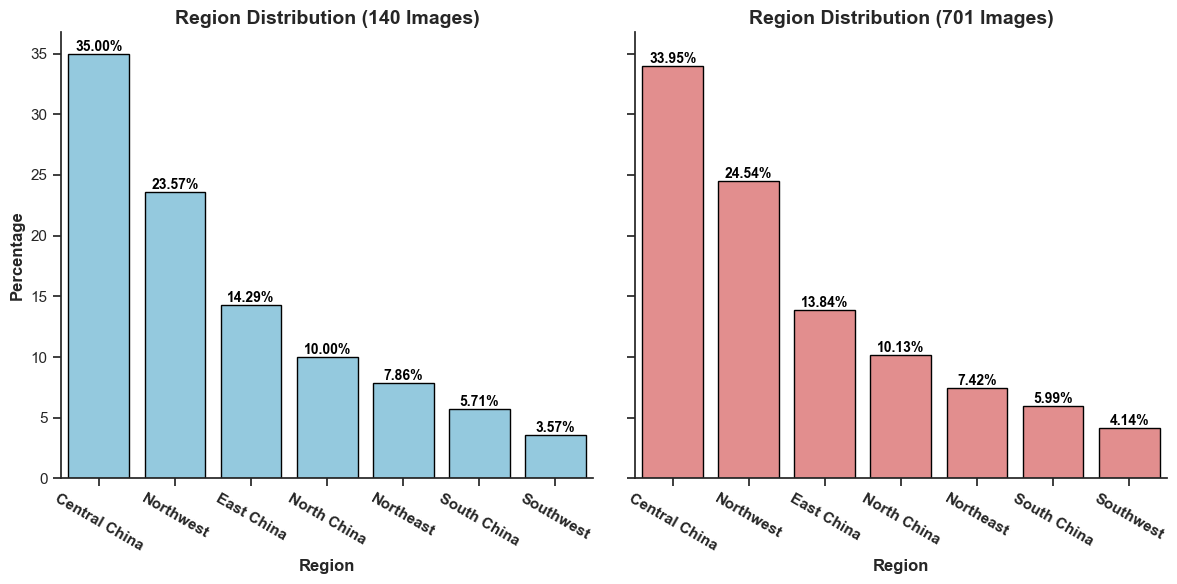}
\caption{\label{figure:sample filter} The human geography region distribution between 140 sampled and total 701 paper-cuttings.}
\Description{The distribution between total 701 and 140 sampled paper-cuttings. This figure compared the human geography region distribution of randomly sampled subset and total 701 papaer-cuttings.}
\end{figure}

\revisedtext{As a regionally influenced art form, we utilized the regional distribution as the standard for the selection of 140 paper-cuttings, as illustrated in~\autoref{figure:sample filter}.}

\subsection{\rrtext{Description of Expert Discussion in the Content Analysis}}
\subsubsection{\rrtext{For the Codebook of Core Factors of Paper-cutting Design Ideation}}\label{A:expert discuss ideation}
\rrtext{The initial version of the codebook was developed solely based on style and function. Thus, in the first round of discussions, the experts recommended adding two dimensions, including subject matter (P1-P2 and P4-P5) and method of expression (P2 and P4-P5), which were accepted through all experts' consensus. During the second round, P2 and P5 suggested merging specific types, such as plants, animals, and mythical creatures into a broader ``flora and fauna'' type. By the third round of discussions, the experts collectively agreed that the current version of the codebook was generally adequate.}

\subsubsection{\rrtext{For the Codebook of Patterns in Paper-cutting}}\label{A:expert discuss pattern}
\rrtext{During the first round of discussions, the experts observed that, in addition to unit geometry and sawtooth patterns, many \textit{Unit Patterns}, such as crescent, cloud, and fire patterns, possess more practical semantics rather than purely abstract ones. As a result, they recommended introducing a new sub-category within \textit{Unit Patterns}, termed unit semantic patterns. Regarding \textit{Composite Patterns}, the experts suggested distinguishing a type that emphasizes functionality and conveys core themes, referred to as composite primary patterns, as opposed to decorative patterns. Following the second round of discussions, the experts collectively agreed that the current version of the codebook was generally appropriate.}

\subsection{\revisedtext{Coding Results of the Ideation Factors of Paper-cuttings}}\label{A:paper-cut coding}
\begin{figure}[H]
\centering
\includegraphics[width=0.49\textwidth]{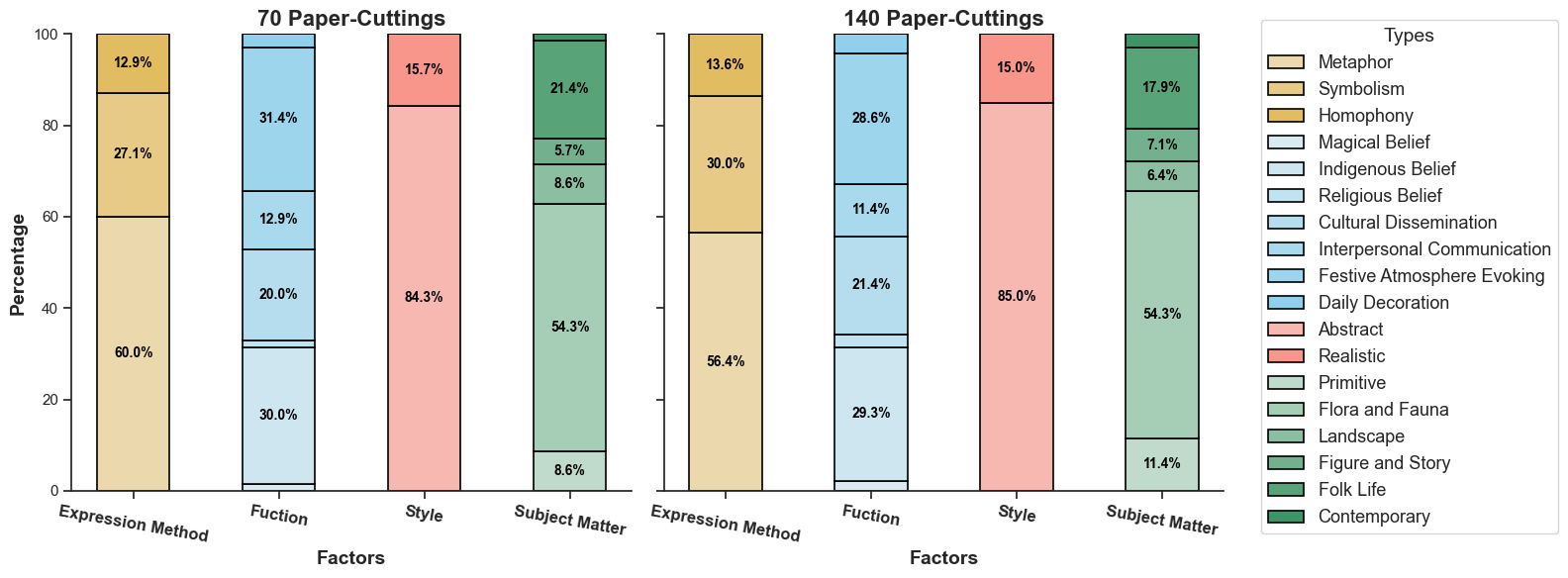}
\caption{\label{figure:paper-cut validation}The coding distribution results of Factors and Types for 70 selections and results in 140 paper-cuttings. Types with less than 5\% are not presented with specific percentages. The differences between the corresponding Types are less than 5\% for all Factors, indicating a similar distribution across both datasets.}
\Description{This figure shows coding distribution results of Factors and Types for 70 selections and results in 140 paper-cuttings, which validate the consistency of coding from the construction to the result.}
\end{figure}

\revisedtext{The first author validated that the taxonomy of ideation factors derived from the analysis of 70 paper-cuttings represented the entire set of 140 paper-cuttings in~\autoref{figure:paper-cut validation}.}

\subsection{Example of Paper-cuttings with Different Type under Design Factors}\label{A:content examples}
\begin{figure}[H]
\centering
\includegraphics[width=0.49\textwidth]{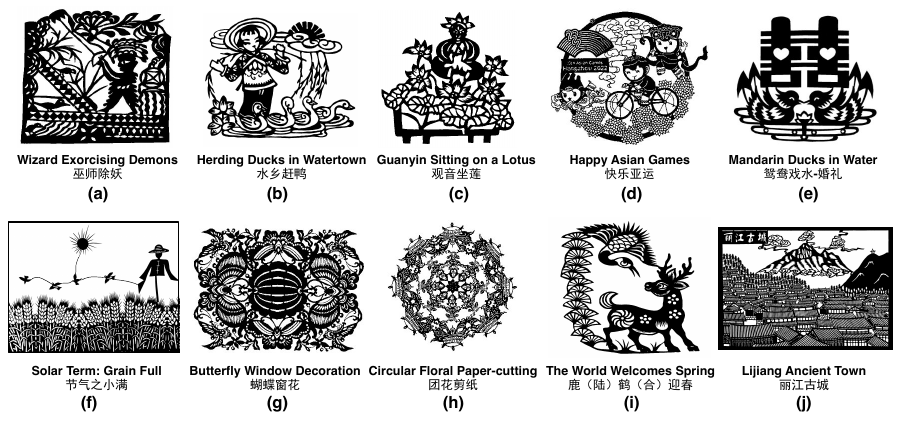}
\caption{\label{a1fig1} Paper-cutting examples that meet design factors. (a) Witchcraft Belief; (b) Indigenous Belief; (c) Religious Belief; (d) Cultural Dissemination; (e) Interpersonal Communication; (f) Festive Atmosphere Evoking; (g) Daily Decoration; (h) Primitive Paper-cutting;  (i) Flora and Fauna; (j) Landscape.}
\Description{This figure shows the paper-cutting examples that meet design factors. (a) Witchcraft Belief; (b) Indigenous Belief; (c) Religious Belief; (d) Cultural Dissemination; (e) Interpersonal Communication; (f) Festive Atmosphere Evoking; (g) Daily Decoration; (h) Primitive Paper-cutting;  (i) Flora and Fauna; (j) Landscape.}
\end{figure}

\begin{figure}[H]
\centering
\includegraphics[width=0.49\textwidth]{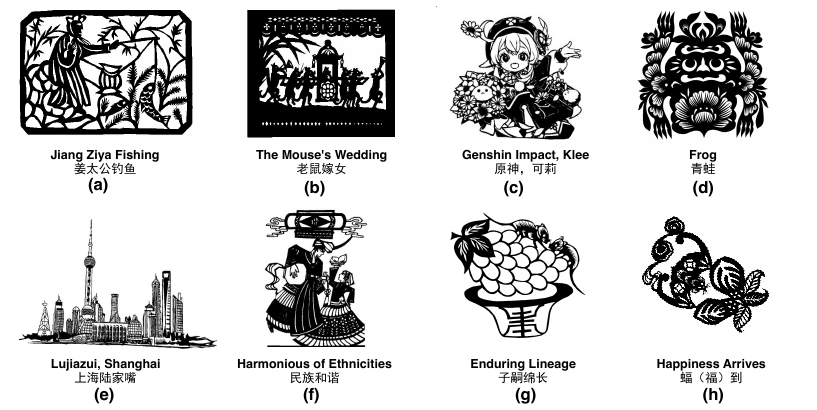}
\caption{\label{a1fig2} Paper-cutting examples that meet design factors (a) Historical Figure and Story; (b) Folk Life; (c) Contemporary Subject; (d) Abstract Style; (e) Realistic Style; (f) Metaphor; (g) Symbolism; (h) Homophony.}
\Description{This figure shows the paper-cutting examples that meet design factors (a) Historical Figure and Story; (b) Folk Life; (c) Contemporary Subject; (d) Abstract Style; (e) Realistic Style; (f) Metaphor; (g) Symbolism; (h) Homophony.}
\end{figure}

\revisedtext{A sequential presentation of paper-cutting examples is provided in~\autoref{a1fig1} and~\autoref{a1fig2}, with each example illustrating the corresponding ideation factor types.}

\subsection{\revisedtext{Coding Results of the Patterns of Paper-cuttings}}\label{A:pattern coding}
\begin{figure}[H]
\centering
\includegraphics[width=0.49\textwidth]{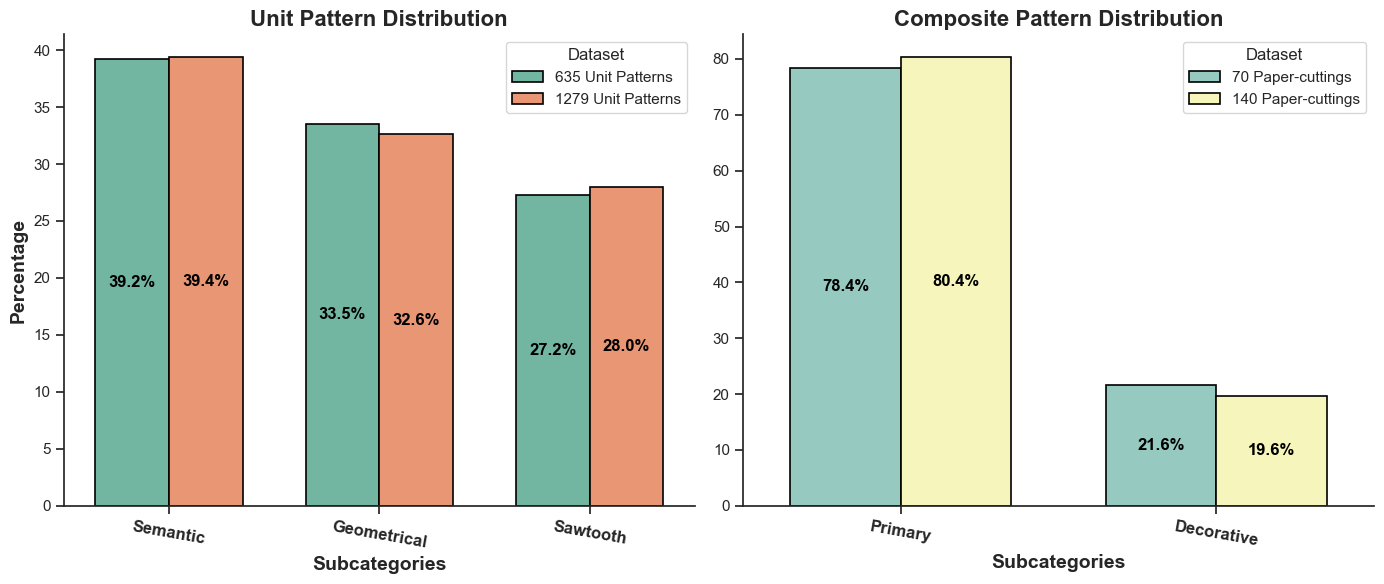}
\caption{\label{figure:pattern validation}The coding distribution results: 635 selections and 1269 cut-outs for the Unit Pattern, 70 selections and 140 paper-cuttings for the Composite Pattern. Each sub-category with less than 5\% is not presented with specific percentages. The differences between the corresponding sub-categories are less than 5\% for all Factors, indicating a similar distribution across both datasets.}
\Description{This figure shows coding distribution results. The coding distribution results: 635 selections and 1269 cut-outs for the Unit Pattern, 70 selections and 140 paper-cuttings for the Composite Pattern, which validate the consistency of coding from the construction to the result.}
\end{figure}

\revisedtext{The first author validated the taxonomy of patterns, including the \textit{Unit Pattern}, derived from the analysis of 635 cut-outs representing the entire set of 1269 cut-outs, and the \textit{Composite Pattern}, derived from the analysis of all composite patterns in 70 paper-cuttings, representing the entire set of 140 paper-cuttings, as shown in~\autoref{figure:pattern validation}.}

\section{\rrtext{Evaluation}}
\begin{table}[H]
\caption{\rrtext{Summary of participants in the user study and the expert interview.}}
  \Description{This table demonstrates the summary of participants in the user study and the expert interview.}
  \label{table:evaluation participants}
\resizebox{0.49\textwidth}{!}{
\renewcommand\arraystretch{1.4}
\begin{tabular}{ccccccc}
\hline
ID  & Sex    & Age & Paper-cutting Expertise & GenAI Expertise    & Location      & Platform \\ \hline
P1  & Male   & 22  & Novice                  & Knowledgeable User & North China   & WeChat   \\
P2  & Female & 28  & Practitioner (18 years) & Knowledgeable User & East China    & Bilibili \\
P3  & Male   & 24  & Novice                  & Knowledgeable User & East China    & WeChat   \\
P4  & Male   & 22  & Novice                  & Knowledgeable User & East China    & WeChat   \\
P5  & Male   & 26  & Amateur (3 years)       & Knowledgeable User & Central China & WeChat   \\
P6  & Female & 25  & Novice                  & Knowledgeable User & Southwest     & WeChat   \\
P7  & Male   & 49  & Master (40 years)       & Novice             & Northwest     & Douyin   \\
P8  & Female & 24  & Practitioner (18 years) & Novice             & Southwest     & Bilibili \\
P9  & Male   & 19  & Novice                  & Knowledgeable User & Southwest     & WeChat   \\
P10 & Male   & 19  & Novice                  & Knowledgeable User & South China   & WeChat   \\
P11 & Male   & 23  & Novice                  & Novice             & East China    & WeChat   \\
P12 & Male   & 18  & Novice                  & Novice             & North China   & WeChat   \\
P13 & Female & 24  & Novice                  & Knowledgeable User & South China   & WeChat   \\
P14 & Female & 18  & Amateur (1 year)        & Knowledgeable User & South China   & WeChat   \\
P15 & Female & 19  & Amateur (2 years)       & Knowledgeable User & Central China & WeChat   \\
P16 & Male   & 23  & Novice                  & Knowledgeable User & Southwest     & WeChat   \\ \hline
E1  & Male   & 30  & Master (21 years)       & Novice             & Central China & Bilibili \\
E2  & Female & 59  & Master (40+ years)      & Novice             & Northeast     & Douyin   \\
E3  & Female & 40  & Master (24 years)       & Novice             & Central China & Douyin   \\ \hline
\end{tabular}
}
\end{table}

\end{document}